\documentclass[acmtog]{acmart}
\usepackage{booktabs} % For formal tables
\usepackage{enumitem}
\usepackage{color,soul}
\usepackage{xcolor}
\usepackage{algorithm}
\usepackage{algpseudocode}
\usepackage{makecell}

\newcommand{\MYhref}[3][blue]{\href{#2}{\color{#1}{#3}}}

\acmSubmissionID{422s1}
\usepackage{color}
\usepackage{bm}

\usepackage{caption}
\usepackage{subcaption}
\usepackage{amsmath,amsbsy}
\usepackage[binary-units]{siunitx}
\usepackage{array}
\usepackage{multirow}
\usepackage{mathbbol}

\usepackage{mathrsfs}   %The amsmath package is included for \xrightarrow

\newcommand{\avof}[1]{\bar{#1}}

\newcommand{\Fobsaw}{\avof{\mathbf{Y}}}

\definecolor{zekun_color}{rgb}{0,0.5,1}
\definecolor{serge_color}{rgb}{0.2,.64,0}
\definecolor{abe_color}{rgb}{1,0,1}
\definecolor{peter_color}{rgb}{0,0.5,0}
\iftrue
\newcommand{\serge}[1]{\textsf{\textcolor{serge_color}{[{\bf Serge}: #1]}}}
\newcommand{\abe}[1]{\textsf{\textcolor{abe_color}{[{\bf Abe}: #1]}}}
\newcommand{\peter}[1]{\textsf{\textcolor{peter_color}{[{\bf Peter}: #1]}}}
\else
\newcommand{\zekun}[1]{}
\newcommand{\serge}[1]{}
\newcommand{\abe}[1]{}
\newcommand{\peter}[1]{}
\fi

\newcommand{\beq}{\begin{equation}}
\newcommand{\eeq}{\end{equation}}

\newcommand{\bal}{\begin{align}}
\newcommand{\eal}{\end{align}}

\newcommand{\obs}{y}

\newcommand{\obsa}{\avof{y}}

\newcommand{\code}{c}
\newcommand{\codei}[1]{\code{}_{#1}}

\newcommand{\noise}{n}
\newcommand{\noisev}{\mathbf{\noise}}

\newcommand{\window}{w}
\newcommand{\obsw}{\mathbf{\obs}}
\newcommand{\obsaw}{\mathbf{\obsa}}

\newcommand{\codev}{\mathbf{\code}}
\newcommand{\codew}{\codev}

\newcommand{\codewd}{\codev{}}

\newcommand{\codewi}[1]{\codev_{#1}}

\newcommand{\codewid}[1]{\codev_{#1}}

\newcommand{\coded}{\codev{}'}

\newcommand{\codedd}{\coded{}}

\newcommand{\codedi}[1]{\codev{}_{#1}'}

\newcommand{\codedid}[1]{\codev{}_{#1}'}
\newcommand{\noisew}{\noisev{}}

\DeclareMathOperator*{\argmax}{argmax}

\newcommand{\nci}{NCI}

\newcommand{\pix}{x}

\newcommand{\FCodew}{\mathbf{C}}

\newcommand{\speedFactor}{\rho}
\newcommand{\speedFactorCandidate}{\speedFactor{}'}

\newcommand*\Bell{\ensuremath{\boldsymbol\ell}}

\newcommand{\reflectance}{r}

\newcommand{\reflectancei}[1]{\reflectance{}_{#1}}
\newcommand{\motionweight}{g}
\newcommand{\motionweightt}{g}

\newcommand{\motionweightatxt}{\mathbf{\motionweight{}}(\pix,t)}
\newcommand{\reflectancev}{\mathbf{\reflectance}}

\newcommand{\codeimagei}[1]{\reflectance_{#1}}

\newcommand{\motiontransportw}{\hat{\reflectance{}}}
\newcommand{\motiontransportiat}[1]{\motiontransportw_{#1}(\pix,t)}

\newcommand{\transporti}[1]{\reflectance{}_{#1}}
\newcommand{\transportiat}[1]{\transporti{#1}(\pix)}

\newcommand{\light}{l}
\newcommand{\lightv}{\Bell{}}
\newcommand{\lighti}[1]{\light{}_{#1}}

\newcommand{\lightw}{\lightv{}}

\newcommand{\Fig}{Figure}
\newcommand{\Figs}{Figures}
\newcommand{\Eq}{Eq.}
\newcommand{\Eqr}[1]{Eq. \eqref{#1}}
\newcommand{\Sectn}[1]{Section #1}

\newcommand{\FigRef}[1]{\Fig{} \ref{#1}}

\newcommand{\noisesum}{\hat{\mathbf{n}}}

\newcommand{\noisesigma}{$\sigma^2$}

\newcommand{\zekun}[1]{{\textcolor{blue}{[Zekun: #1]}}}

\newcommand{\E}{\mathbb{E}_{\noisew{}}}

\newcommand{\bulletitem}[1]{\item\textbf{#1}}

\newcommand{\bracetext}[1]{\text{\footnotesize #1}}

\definecolor{mathbrace_color}{rgb}{0.2,0.5,1.0}
\newcommand{\cunderbrace}[2]{\color{mathbrace_color}\underbrace{\color{black}#1}_{\bracetext{#2}}\color{black}}
\newcommand{\cunderbracelines}[3]{\color{mathbrace_color}\underbrace{\color{black}#1}_{\substack{\bracetext{#2}\\\bracetext{#3}}}\color{black}}
\newcommand{\coverbrace}[2]{\color{mathbrace_color}\overbrace{\color{black}#1}^{\bracetext{#2}}\color{black}}
\newcommand{\coverbracelines}[3]{\color{mathbrace_color}\overbrace{\color{black}#1}^{\substack{\text{\scriptsize#2}\\\text{\scriptsize#3}}}\color{black}}

\newcommand{\alignedAlignmentTerm}{\codew{}^\intercal\codew{}\reflectance{}}
\newcommand{\alignmentTerm}{\codedd{}^\intercal\codew{}\reflectance{}}
\newcommand{\spatial}{M}
\newcommand{\V}{Std_\noisew{}}
\newcommand{\Rms}{Rms}

\newcommand{\stdn}{\sigma_n(L)}

\newcommand{\subbullet}{$\circ{}$}

\newcommand{\LConstw}{\mathbf{L}}

\newcommand{\bilateral}{\mathcal{B}}
\newcommand{\bilateralof}[1]{\bilateral{}(#1)}

\newcommand{\alignmentcurve}{alignment curve}

\newcommand{\AMatrices}{Alignment Matrices}

\newcommand{\amatrices}{alignment matrices}
\newcommand{\amatrix}{alignment matrix}

\newcommand{\transientfiltered}{transient-filtered}

\newcommand{\TransientFiltered}{Transient-Filtered}

\newcommand{\nscenes}{9}

\newcommand{\toffset}{\delta{}}

\newwrite\graphics
\immediate\openout\graphics=\jobname.graphics%
\let\oincludegraphics\includegraphics% store original \includegraphics
\renewcommand{\includegraphics}[2][]{% prepend to it (could also use xpatch, etc.)
\immediate\write\graphics{#2}
\oincludegraphics[#1]{#2}}

\newcommand{\codeshadows}{code shadows}

\usepackage[commandnameprefix=ifneeded]{changes}
\usepackage{marginnote}

\newif\ifshowchanges
\showchangesfalse

\definecolor{forestgreen}{RGB}{0, 120, 0}
\definecolor{cerulean}{RGB}{42, 82, 190}
\definecolor{blue}{RGB}{0, 0, 220}

\definecolor{abe_addcolor}{rgb}{0,0.3,0.8}
\definecolor{abe_removecolor}{rgb}{0.9,0.2,0.1}

\definecolor{petercolor}{rgb}{0.7,0.2,0.7}
\definecolor{peter_addcolor}{rgb}{0.7,0.2,0.7}
\definecolor{peter_removecolor}{rgb}{0.9,0.6,0.9}

\newcommand{\PeterID}{PM}

\definecolor{marginotecolor}{RGB}{100,175,255}

\ifshowchanges
    \newcommand{\aadd}[1]{{\leavevmode\textcolor{forestgreen}{{#1}}}}
    
    \newcommand{\adelete}[1]{{\leavevmode\textcolor{abe_removecolor}{{\sout{DeletedPart}}}}}
    
    \newcommand{\acomment}[1]{\comment[id={\AbeD}]{#1}}

    \newcommand{\padd}[1]{{\leavevmode\textcolor{forestgreen}{{#1}}}}
    
    \newcommand{\pdelete}[1]{{\leavevmode\textcolor{peter_removecolor}{{\sout{#1}}}}}
    
    \newcommand{\pcomment}[1]{\comment[id={\PeterID{}}]{#1}}

    \newcommand{\revcomment}[1]{\marginpar{\par\noindent\colorbox{marginotecolor}
{\parbox{1cm}{\footnotesize #1}}}}
   
    \newcommand{\changenote}[2][]{\added[#1,comment={#2}]{}}
\else
    \newcommand{\aadd}[1]{#1}
    
    \newcommand{\adelete}[1]{}
    
    \newcommand{\acomment}[1]{}

    \newcommand{\padd}[1]{#1}
    
    \newcommand{\pdelete}[1]{}
    
    \newcommand{\pcomment}[1]{}

    \newcommand{\edelete}[1]{}
    
    \newcommand{\ecomment}[1]{}

    \newcommand{\changenote}[2][]{}
    \newcommand{\revcomment}[1]{}
    
\fi

\begin{document}

% Title.
% If your title is long, consider \title[short title]{full title} - "short title" will be used for running heads.
\title{Noise-Coded Illumination}
\subtitle{For Forensic And Photometric Video Analysis}

\author{Peter F. Michael}
\affiliation{
  \institution{Cornell University}
  \country{USA}
}
\author{Zekun Hao}
\affiliation{
  \institution{Cornell Tech}
  \country{USA}
}
\author{Serge Belongie}
\affiliation{
  \institution{University of Copenhagen}
  \country{DK}
}
\author{Abe Davis}
\affiliation{
  \institution{Cornell University}
  \country{USA}
}

% This command defines the author string for running heads.
% \renewcommand{\shortauthors}{DeJohnette, Rowland-Smith, Badeeri, and Foyt}
\renewcommand{\shortauthors}{P.F. Michael et al.}
\setcounter{secnumdepth}{4}

% abstract
\begin{abstract}

The proliferation of advanced tools for manipulating video has led to an
arms race, pitting those who wish to sow disinformation against those who
want to detect and expose it. Unfortunately, time favors the ill-intentioned in this race, with fake videos growing increasingly difficult to distinguish from real ones. 
At the root of this trend is a fundamental advantage held by those manipulating media: equal access to a distribution of what we consider
authentic (i.e., ``natural'') video. 
In this paper, we show how coding very subtle, noise-like modulations into the illumination of a scene can help combat this advantage by creating an information asymmetry that favors verification. Our approach effectively adds a temporal watermark to any video recorded under coded illumination. However, rather than encoding a specific message, this watermark encodes an image of the unmanipulated scene as it would appear lit only by the coded illumination. We show that even when an adversary knows that our technique is being used, creating a plausible coded fake video amounts to solving a second, more difficult version of the original adversarial content creation problem at an information disadvantage. This is a promising avenue for protecting high-stakes settings like public events and interviews, where the content on display is a likely target for manipulation, and while the illumination can be controlled, the cameras capturing video cannot.

\end{abstract}

%CCS


%keywords

% A "teaser" figure, centered below the title and authors and above the body of the work.
\begin{teaserfigure}
  \centering
  \includegraphics[width=\textwidth]{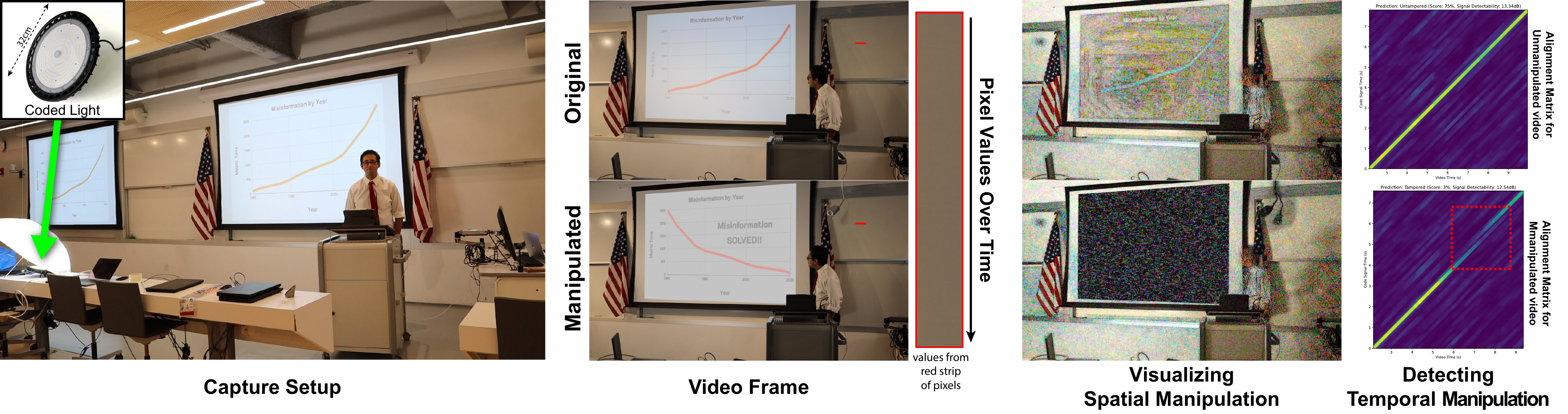}
  \caption{
  % We present \textbf{\emph{Noise-Coded Illumination} (\nci{})}, a technique for watermarking light in a physical environment that makes spatial and temporal manipulation of a video much simpler to detect. 
  We present \textbf{\emph{Noise-Coded Illumination} (\nci{})}, a technique for watermarking light that makes it much easier to detect spatial and temporal manipulation of video captured in a specific environment.
  The left shows a scene with one coded light placed in a lecture hall during a presentation. The top middle image shows a frame from an unaltered video of the scene. The bottom middle image shows a frame from a manipulated video, where the presentation slide content has changed, a surveillance camera has been added to the back wall using a generative model, and part of the recorded dialogue has been removed with a warp cut to change its meaning. The vertical strip outlined in red visualizes color values taken from a small strip of pixels (highlighted in red) over time. The strip looks mostly constant, reflecting the subtlety of coded light variations. The subsequent column of images ("Visualizing Spatial Manipulation") contain code images recovered from the original video (top) and the manipulated video (bottom). Unmanipulated code images should approximate what the scene would look like if it were illuminated only by the coded light placed to the left of the scene. We see that fake content does not appear to reflect light in the bottom code image extracted from the manipulated video, indicating that it was not part of the original scene.
  % light in the code image on the bottom, indicating that they are  we see that the fake content added to the manipulated video appears black in its recovered code image. 
  The plots of the rightmost column show temporal \amatrices{} extracted from each video. The \amatrix{} for the original video shows an uninterrupted diagonal, while the matrix for the manipulated video shows a discontinuity where a word was removed (highlighted by the red box). Please see our \MYhref{http://peterfmichael.com/nci}{website} for video results.}
  \Description{Left To Right: image of capture setup. Top: frame from unaltered video. Bottom: frame from manipulated video. Strip of pixel values over time to illustrate subtlety of approach. Top: code image from unaltered video. Bottom: code image from manipulated video. Top: temporal \amatrices{} for unmanipulated video. Bottom: temporal \amatrices{} for manipulated video.}
  \label{fig:teaser}
\end{teaserfigure}

\acmJournal{TOG}
% \acmVolume{42}
% \acmNumber{3}
% \acmYear{2025}
% \acmMonth{6}
\maketitle
\section{Introduction}
The past decade has seen incredible progress in the technologies used to capture, edit, and share video. 
% This progress proved especially valuable during the pandemic, as much of the world became more reliant on video-based communication to stay safe and connected. 
% But as video has taken on an increasingly important role in how we communicate increasingly difficult to avoid recording and sharing videos on a regular basis---deepening fears that such videos could be re-purposed to create and spread misinformation. 
%%
At the same time, the proliferation of increasingly advanced, low-cost editing tools has made manipulating video easier than ever before, deepening fears that edited content may be used to spread disinformation. The malicious use of such content has evolved from a hypothetical threat into a real and present problem, with doctored video spreading disinformation to millions of users on social media in recent years.\footnote{Examples include the incidents involving Jim Acosta \shortcite{AcostaFake}, Nancy Pelosi \shortcite{PelosiSlur}, and Joe Biden \shortcite{BidenShallowFake}, which we examine later in this paper.} Unfortunately, this problem has only gotten worse over time, as the progress of video editing tools has far outpaced that of forensic techniques. This trend points to an urgent need for new techniques that can create an information advantage for forensic analysis.
One promising strategy is to digitally watermark content when it is first encoded by a camera \cite{watermarknoise}. This approach has many advantages, but requires access to the capturing camera, which is impractical in common scenarios where video may be recorded by uncontrolled third parties. We propose a new type of watermarking, which we call \emph{noise-coded illumination} (\nci{}), that instead watermarks the illumination in a scene. 
Our approach
% , which we call \emph{Noise-Coded Illumination} (\nci{}), 
works by modulating the intensity of each light source by a subtle pseudo-random pattern drawn from a distribution that resembles existing noise. To observers, video captured under \nci{} is at best indistinguishable from regular video, and at worst contains a subtle flicker that resembles some fluorescent and LED lighting. However, hidden in the apparent noise of these videos is a different \emph{code image} for each coded light source present in the scene. When an adversary manipulates video captured under coded illumination, they unwittingly change the code images contained therein. Knowing the codes used by each light source lets us recover and examine these code images, which we can use to identify and visualize manipulation. 
\subsection{Design Goals}
\label{sec:designgoals}
There is no perfect solution to the video forensics problem, but adding new forensic tools can make it more difficult for an adversary to circumvent all of them. As such, our technique does not need to work in every scenario for it to be useful, but it should offer protections that complement or extend existing approaches. With this in mind, \nci{} distinguishes itself from other techniques by balancing the following goals:

\begin{enumerate}[leftmargin=*]
\item \textbf{Information Asymmetry:}\;
One of the biggest challenges facing video forensics is editing tools that can learn from massive amounts of data. When such tools are sufficiently trained, their outputs can be near-indistinguishable from authentic video. Information asymmetry means that the analyst should have some extra information that is not available to the adversary and not learnable from publicly available training data.
% combats this by ensuring that the analyst has some extra information about the subject of video that would not be reflected in possible training data.

% that is not present in training data that an adversary would have access to.
% For example, with digital watermarking, the analyst knows that authentic video should contain some hidden watermark. 
% As a type of watermarking, \nci{} combats this by shifting the distribution of plausible video, and knowledge of the code used by each light source provides an information advantage when estimating this shift (see Section \ref{sec:theory}).
% lets one predict this shift, offering an information advantage when authenticating video. Furthermore, while an informed adversary (one that knows \nci{} is being used) may try to estimate the code, this can only be done up to some limited SNR, leaving a smaller but still significant information advantage.
\item\textbf{Interpretability:}\;
% There is a fundamental tradeoff between the certainty that a forensic tool offers and the conditions it requires for use. 
There are many ways that a video can be modified, and not all of them carry the same meaning, which makes it important to distinguish between different types of manipulation. For example, a simple checksum can tell if a video file has been changed, but will not differentiate between simple video compression and more potentially dangerous changes like the insertion of virtual objects. Interpretability means that we can interpret and distinguish between different types of manipulation, which is particularly important for identifying disinformation.
% There are many ways that a video can be modified, and not all of them carry the same meaning, which makes it important to distinguish between different types of manipulation. For example, a simple checksum can tell if a video file has been changed, but will not differentiate between simple video compression and more potentially dangerous changes like the insertion of virtual objects. Interpretability means the ability to identify what type of manipulation was performed, and it is one of the most significant strengths of our technique.
% Another goal of \nci{} is to help analysts interpret and distinguish between different types of manipulation.
% Compared with most watermarking techniques \nci{} is highly interpretable and can be used to distinguish between many different types of manipulation. This is in part because, as an imaging method, \nci{} complements other forensic strategies based on natural image or video priors. 
\item\textbf{Indirect Application:}\;
% Digital watermarking creates information asymmetry by directly modifying captured video. 
One way to create information asymmetry is by directly modifying captured video in some secret way. Unfortunately, this requires control over the capturing camera, which is impossible in many real-world settings. Indirect application means that we do not require any control over the recording camera.
% \nci{} addresses this limitation by applying a physical watermark to the light that illuminates a subject.
\item\textbf{Subtlety:}\;
Our approach should not obstruct the scene or place any additional burdens on those using the protected space. An uninformed observer should not even notice that the method is being used.
\item\textbf{Reference-Free:}\;
Our approach should not require access to any original, unmanipulated video. It should work even in cases where the only video available of an event has been manipulated.
\end{enumerate}

\noindent While some other watermarking techniques achieve information asymmetry, they generally fall short on one or more of our remaining goals. For example, digital watermarking requires direct access to the camera, and the more recent approach of \cite{totems} requires that the capturing camera keep multiple glass orbs in frame during capture. \nci{} achieves our indirect application and reference-free goals by hiding its watermark in the apparent noise of illumination. Subtlety and interpretability are accomplished by careful design and analysis of our watermark, which we discuss in Sections \ref{sec:theory}-\ref{sec:mouncoded}.

% Notably, most methods that  these goals can be difficult. For example, digital watermarking create information asymmetry but Information asymmetry is simple to achieve with digital watermarking, but this makes the remaining tht

% Most existing forensic tools can be classified as either digital watermarking or analysis based on natural video priors. Some digital watermarking techniques already satisfy our first goal of information asymmetry, but not the remaining two goals. Meanwhile, techniques based on natural video priors often satisfy the second two goals, but not the first. A key strength of \nci{} comes from combining these two broader approaches, which is accomplished by coding information analogous to images of the unmanipulated scene into a physical watermark.

\subsection{Scope \& Methodology}
As with any forensic tool, \nci{} should always be applied as part of a broader analysis that takes into consideration the context and circumstances of a video. With this in mind, our results focus primarily on visualization and quantitative evaluations that serve this type of analysis, which we apply to a broad range of experiments examining:

\begin{itemize}[leftmargin=10pt]
\item A variety of manipulations that have been used in recent examples of disinformation, including:
\begin{itemize}
    \item[\subbullet{}] Temporal edits, including warp cuts, as well as speed, and acceleration manipulation
    \item[\subbullet{}] Spatial edits, including compositing and deep fakes
\end{itemize}
\item Robustness to various factors including: 
\begin{itemize}
    \item[\subbullet{}] Different signal levels, including those near and below the human perception threshold
    \item[\subbullet{}] Subject and camera motion
    \item[\subbullet{}] Transient phenomena like camera flash
    \item[\subbullet{}] Different levels of video compression
    \item[\subbullet{}] Human subjects with different skin tones
    \item[\subbullet{}] Indoor and outdoor settings
\end{itemize}
\end{itemize}

\noindent Additionally, while it is impossible to anticipate every attack an adversary might use, we examine robustness to basic attacks from an informed adversary (i.e., one that knows \nci{} is being used) based on both local and global strategies for estimating illumination codes \aadd{(see experiments in our supplemental material)}. 
% Our experiments are extensive for work that introduces a new approach,
% introduction to this new approach, and they show a great deal of promise for \nci{}, 
% but 
This work is only the first step in a new research direction. \Sectn{\ref{sec:limit}} discusses current limitations of the approach, and \Sectn{\ref{sec:futurework}} discusses possible future directions and improvements.

\begin{figure*}
    \centering
    \includegraphics[width=\textwidth]{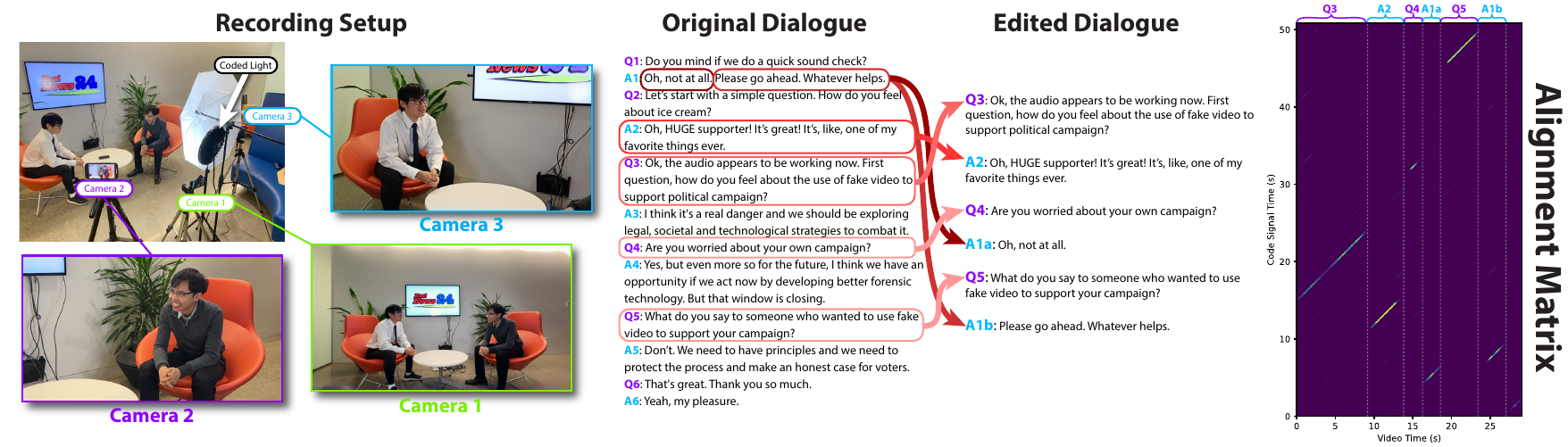}
    \caption{\textbf{Characterizing Misleading Cuts} Our interview scene follows a standard three-camera television interview setup (left) consisting of one wide-angle camera (Camera 1) one close-up camera for the interviewer (Camera 2) and one for the interviewee (Camera 3). In the original dialogue of our scene, the interviewee expresses concern about fake video in political campaigns (mid left). The maliciously edited version splices in footage from an earlier response given during a sound check that makes it look like the interviewee supports and encourages the use of fake video for spreading disinformation (mid right). Our recovered alignment matrix displays the original timing of each clip, which shows that the answers in the manipulated video came from footage recorded before the corresponding questions, indicating that they were maliciously taken out of context.}
    \Description{Left to Right: image of capture setup. Original and edited dialogue. Alignment matrix of manipulated clip.}
    \label{fig:interview}
\end{figure*}

\section{Related Work}
% Our work takes inspiration from several sources, building on previous approaches to video forensics as well as theory from wireless communications and techniques in computational illumination and photography.

\subsection{Image and Video Forensics}

Most work on video forensics falls into one of two categories, which differ in their assumptions.

% The field of image and video forensics has been explored from many angles but is fast-evolving by its nature. We broadly categorize these techniques into two categories: passive and active.

\subsubsection{Passive / Natural Prior-Based Techniques}
The most general form of the video forensics problem is to determine whether a video is authentic using only natural video priors. These are passive in that they do not assume any control over the capturing camera or its subject.
% Passive techniques for image and video authentication do not assume any special prior on the captured data. 
\cite{faridreflections, faridshadows, faridshading} detect tampering done to images by looking at inconsistencies in object reflections, shadows, and shading.
Some passive approaches look for anomalies in the noise characteristics of an image \cite{lukavs2006detecting} or video \cite{Verdoliva_2019_CVPR_Workshops}. MP4 Tree Networks (MTNs) \cite{mtn} learn to detect manipulations from the statistics of MP4 files instead of decoded pixel data.
% It uses a graph neural network to learn a representation of this structure for manipulation detection. 
Other works look for statistical anomalies in audiovisual data. \cite{audvid} takes a self-supervised approach to learning how to detect such anomalies.
% Changing the video or audio of a person changes their identity, thus changing their embedding in this space and providing evidence for manipulation when compared to their original identity embedding. 
\cite{freq} detect deep fakes by examining the spatial frequency content of manipulated images and video. These approaches are limited by a lack of information asymmetry: in the long term, adversaries can learn and exploit the same natural image priors to circumvent their use.  
% These approaches 
% While these approaches can work in specific cases, they are unreliable in the long term since they exploit shortcomings in current approaches that will become increasingly difficult to find.
% We discuss limits of passive techniques in Section \ref{sec:principles}.

\subsubsection{Watermarking \& Steganography}
Techniques based on watermarking involve some detectable modification of authentic content when it is first created. Watermarking was originally an anti-counterfeiting measure used on printed documents. Some distinct mark would be applied to the original document that could be used later for verification. Analogous strategies have been used on images and video, mostly for copyright protection or verifying the provenance of image content \cite{videowatermark, imagewatermark,watermarkreview}. Modern approaches typically use unique and imperceptible watermarks, which can also be considered a form of steganography (the practice of hiding information in other content to avoid detection). Recent approaches have used neural networks to embed imperceptible watermarks in images or video \cite{neuralimagingpipeline, rosteal, proactive, cmua, facesigns}. Most similar to our work in this space is the approach of \cite{watermarknoise}, which injects near imperceptible pseudo-random noise in a video as a post-process. While most of these techniques create information asymmetry, they generally do not satisfy our goals of indirect application or interpretability. \nci{} addresses this weakness by watermarking the light in a physical environment.

\subsubsection{Content-Specific Priors}
% \cite{avpoi} uses contrastive learning to find audiovisual embeddings for individual human subjects, which can be used to detect manipulations that are inconsistent with a given person's appearance and behavior. 
Some forensic techniques rely on more content-specific priors. For example, training a prior on a particular individual's behavior \cite{avpoi,agarwal2020detecting}, or on statistical signs that an image is the result of compositing \cite{wang2019detecting, huh2018fighting}.
Work on totems \cite{totems} places glass spheres in the scene and checks for consistency with different views of refracted content. Like \nci{}, this modifies a scene to make manipulation easier to detect. However, their totems must be visible along with the subject, and typically take a substantial portion of the field of view. By contrast, \nci{} has little or no noticeable impact on protected content.

\subsection{Computational Illumination}
Computational illumination has been actively explored for a variety of different applications. Structured light systems, such as those used for face identification on many modern phones, project known illumination patterns onto surfaces to triangulate depth \cite{structlight, kalstructuredlight, unstructuredlight}. Light stages place a subject in a sphere of LEDs to compute bases for relighting \cite{lightstage}. \cite{fnf} and \cite{flashnoflash} uses flash-no flash pairs to denoise images in dark environments. \cite{bounceflash} uses computational techniques to determine the best directions to point a camera flash. Many recent works focus on using more affordable hardware, such as \cite{lightdesk}, which relights an image of a person's face based on observed time-varying illumination from watching content on a computer monitor. \cite{MultiSpectralMultiplexing} and \cite{opticaspectral} use synchronized illumination sources of varying spectra to infer hyperspectral images from RGB or monochrome cameras. Time of flight (ToF) sensing is mostly used for depth estimation \cite{8578766,7298851,SRATOF}, but has also been used for transient imaging \cite{LightInFlightFelix,FDomainTransient,Velten12recoveringthreedimensional,NLOSWOccluders}.

% \marginpar{\marginbox{Hello there this is a margin note}}

\revcomment{Elaborated on differences with these works}
\aadd{%
There are two works that use computational illumination for watermarking or forensics in more limited settings. \cite{spi} explores watermarking for single-pixel camera systems using illumination from a projector, and \cite{farid} focus on real-time protection of video conference calls by using a fixed-area solid region of the screen to display time-varying hue that can be tracked in the video. Neither of these works claim to hide illumination in the scene, detect general or adversarial manipulation, or work in the general settings we explore.
}%

% \comment{Clarified differences and why we do not compare in this work.}
% , Our work shows that we can code arbitrary light sources and use them to detect a wider range of manipulations in more general settings.

% The idea of coding illumination is not new, though most work in this domain codes light outside of the visible spectrum. In particular, time of flight (TOF) imaging has been used to do everything from depth sensing \cite{8578766,7298851,SRATOF} to transient imaging \cite{LightInFlightFelix,FDomainTransient,Velten12recoveringthreedimensional,NLOSWOccluders}. Multiplexed light patterns have been used for multispectral imaging \cite{MultiSpectralMultiplexing} and direct-indirect light separation \cite{FastDirectIndirect}. We also draw inspiration from work in computational photography on multi-image relighting \cite{UserAssistedRelighting}, and coded exposure \cite{CodedShutter,CodedRollingShutter}. The light source separation feature of \nci{} is similar to another work which relights an image of a person's face based on its observed time-varying illumination from watching content such as a YouTube video \cite{lightdesk}.
\subsection{Other Areas}
\subsubsection{Direct-Sequence Spread Spectrum}
One of the most closely-related techniques to our own actually comes from wireless communications. Specifically, direct sequence spread spectrum (DSSS), which is a modulation technique used to spread the transmission of a signal over a broad frequency band by modulating it with pseudo-random noise. DSSS is part of the IEEE 802.11 network specification standard \cite{DSSS_IEEE802p11}. Our illumination code uses a similar principle to minimize human perceptibility and improve robustness to factors like video compression. 
\subsubsection{Subtle Visual Signals}
Another closely related area of work is on analyzing extremely subtle signals in video. In particular, work that has shown how a surprising amount of useful information can be recovered from a temporal analysis of extremely subtle variations in video. This includes motion magnification \cite{Liu:2005,wu2012eulerian,PhaseBasedMM13}, visual vibration analysis \cite{VisMic,DavisISMB,VisVibPAMI}, and imaging the electric grid \cite{ImagingElectricGrid}.

% \subsection{Doctored Video in Politics}

% \cite{UserAssistedRelighting} - photograph relighting by recombining images of a scene under different illuminations.
% \cite{MultiSpectralMultiplexing} - multiplexing colors to recover hyperspectral images.
% \cite{FastDirectIndirect} - projector patterns for direct/indirect separation.
% \cite{ImagingElectricGrid} - imaging the electric grid.
% \cite{CodedRollingShutter} - Coded Rolling Shutter
% \cite{CodedShutter} - Coded exposure photography.

% % strategybecomes much easier when we can rely on a prior that our adversaries (those creating fake content) do not have access to. A classic example of this is \emph{watermarking}. 
% Originally an anti-counterfeiting measure used for printed documents, watermarking works 

% NCI draws inspiration from a variety of work in different domains. Direct Sequence Spread Spectrum (DSSS), a method for spreading the transmission of a signal over a broad frequency band by modulating it with psuedo-random noise. which is part of the IEEE 802.11 network specification standard \cite{DSSS_IEEE802p11}.

% \paragraph{Leveraging a Private Prior}

% Direct Sequence Spread Spectrum (DSSS), a method for spreading the transmission of a signal over a broad frequency band by modulating it with psuedo-random noise. which is part of the IEEE 802.11 network specification standard \cite{DSSS_IEEE802p11}.

% Image forensics: \cite{wang2019detecting} \cite{huh2018fighting}.

% Video forensics: \cite{agarwal2020detecting}
\section{Noise-Coded Illumination}
\label{sec:theory}
% Section \ref{sec:principles} establishes terminology for reasoning about our forensic task. Section \ref{sec:ncitheoryintro} introduces \nci{} in the limited case of 
% This paper introduces \nci{} and performs an extensive analysis of factors that influence its use. This section provides the theoretical motivation for our approach and an overview of topics covered throughout the rest of the paper.

% \abe{bring some of design goals text in here}

% There are several ways to interpret Noise-Coded Imaging: as spatio-temporal watermarking, self-referential steganography, coded light source multiplexing---each of these interpretations can be useful for placing it in context with different related work. Here we focus on an explanation that we found particularly motivating for forensic applications. \abe{Abe change this / update this}
% We begin with high-level intuition dealing with priors on the manifold of all plausible videos, then derive a more explicit formulation for visualizations used throughout the rest of the paper. 

\subsection{The Plausible Video Manifold}
\label{sec:principles}
Forensic video analysis is often discussed in terms of an adversarial game, pitting an analyst against some unknown adversary. 
% The adversary's goal is to produce video that spreads disinformation, while the analyst's goal is to expose that video as fake. 
We can think of this game in terms of a \emph{plausible video manifold}, which represents all video that the analyst cannot identify as untrustworthy. The adversary's goal is then to find a point on the plausible manifold that can be used to spread false or misleading information, and the analyst's goal is to prevent this from happening. In practice, we will often discuss this manifold probabilistically (e.g., how likely is a video on the manifold) to reflect uncertainty, but the manifold concept will help us reason about the impact of different information on our adversarial game.

Without more specific priors on the content of a video,
% more specific information about the content of a video, 
an analyst can only base their plausible manifold on natural video priors. 
Unfortunately, this is a losing strategy in the long term; the space of all natural-looking video includes plenty of content that can be used to convey false information, and as generative models continue to improve, it only becomes easier for adversaries to find this content. Fortunately, additional information about the content of a video can help in two key ways: by reducing the plausible manifold, and by creating information asymmetry. 

% Unfortunately, this is a losing strategy in the long term; as generative models continue to improve, it only becomes easier for adversaries to find natural-looking content that can effectively convey false information.

% In the long term, adversaries and analysts will model these priors equally well, since they have equal access to the data needed to model it. In this case, the adversary wins since the analyst's predictions would be statistically random, representing a long-term disadvantage for those trying to analyze video. 
% Fortunately, additional information about the content of a video can help in two key ways. 
% Here, the challenge for each side depends on what specific knowledge the analyst has about the subject of a given video. 
    
\subsubsection{Reducing the Manifold}
% There is simply no general way to distinguish between video of something that \emph{did} happen and perfectly plausible video of something that \textit{could have} happened. 
% Here is where more specific knowledge about a video can help. For example, c
In general, more specific information about a scene will make the manifold of plausible video smaller. For example, consider a scenario where the analyst has access to the only recording ever made of a particular event (e.g., they can guarantee that no other cameras were present for the event). In this scenario, the analyst can limit the plausible manifold to acceptable variations of that one video (e.g., minor compression or reformatting). This is an extreme example of a prior that reduces the manifold, but an analyst could use less restrictive information to similar effect; for example, the time of an event, weather conditions, or knowledge of people confirmed to be absent from the event. Notably, reducing the manifold is a benefit to the analyst even when an adversary has access to the same information, because a smaller manifold leaves fewer possible vulnerabilities that an adversary could exploit.
% it leaves little opportunity for  content that can be found on the manifold.

\subsubsection{Information Asymmetry}
Information asymmetry helps by making it difficult for an adversary to estimate the plausible manifold, thereby making it more likely that they will push a video off the manifold when manipulating it. 
% This is  which will be easy for the analyst to detect. 
Information asymmetry does not always reduce the manifold: for example, bijective encryption creates information asymmetry without reducing the manifold. 
% A common cost of information asymmetry is the need for some intervention; e.g., digital watermarking requires access to the camera or recorded video. 

\vspace{0.1cm}
Part of what makes \nci{} unique is how it combines information asymmetry with manifold reduction. Like other watermarking techniques, it creates information asymmetry through the use of a secret code. But in the case of \nci{}, the code also acts as a carrier signal for additional information about the scene, which serves to further reduce the plausible manifold. More specifically, the code carries a visual decomposition of illumination in the scene, which makes plausible video more difficult to fake even if an adversary somehow discovers the illumination codes.
This steganographic property also gives us a way to isolate
% This information serves two purposes: first, it makes successfully faking video more difficult, and second, it makes manipulation of a video easier to characterize and interpret.
% This second point is one of the strongest features of \nci{}: it can often isolate 
manipulation in both time and space. For example, we can tell where objects are inserted or removed from video, how much time was removed between cuts, or if clips were sped up, slowed down, or reordered. These details help us reason not just about the presence of manipulation, but also the possible intent behind it. 

 % that is difficult to fake (specifically, light source separation). In other words, our information asymmetry complements a manifold reduction: the code hides extra information about the scene, and that same information makes the code more difficult for an adversary to estimate. 
% this additional information, and the additional information makes the code harder to estimate.
% This additional information makes the code harder for an adversary to estimate, and In this way,  which makes many types of manipulation more difficult even if an adversary somehow discovers the illumination codes.

% , resulting in a coded reduction of the manifold. But unlike digital watermarking techniques, the code in \nci{} is used as a More specifically, it encodes additional information about a scene into 
% \nci{} offers two important protections. First, it creates information asymmetry by using illumination codes that are not known to the adversary. Second, when multiple coded lights are used, it reduces the plausible manifold by encoding a decomposition of scene illumination into the apparent noise of captured video. Neither of these protections is perfect, but they complement each other extremely well. 

\begin{figure}[t]
\begin{center}
    \includegraphics[width=0.9\linewidth]{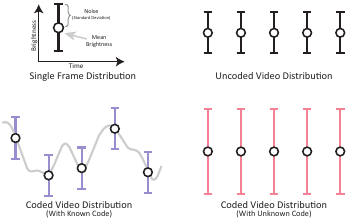}
\end{center}
    \caption{\textbf{Single-Pixel \nci{}:} Consider a single-pixel video of a static scene and camera under constant illumination. A prior for this video (top right) would consist of some mean brightness and zero-mean noise at each frame. If we modulate our illumination by a known code drawn from our noise distribution, then the means in our prior change but the variances do not (bottom left). However, if we do not know the code, variances change and means do not (bottom right). As a result, knowing the illumination code reduces the variance of our prior.} 
\Description{Top Left: vector drawing showing mean and standard deviation of a single pixel video. Top Right: uncoded distribution over time. Bottom Left: coded distribution over time with a known code. Bottom Right: coded distribution over time with an unknown code.}
\label{fig:OneDNCI}
\end{figure}
\subsection{Coding Illumination}
\label{sec:ncitheoryintro}
\nci{} is simplest to understand if we first consider the formation of single-pixel videos in a static scene illuminated by just one light source. In this simplified setting, our video can be modeled as a scalar time signal given in terms of a light transport coefficient
$\reflectance$, the power $\light$ of our light source, and some noise $\noise{}(t)$ drawn from a zero-mean distribution with variance \noisesigma{}: 
\beq
\obs{}(t)=\light{}\reflectance{}+\noise{}(t)
\eeq
\revcomment{Clarified that we handle shot noise and add pointer to supplemental}
\aadd{Our plausible manifold can then be described with a distribution over frame values, where each frame has mean $\light{}\reflectance$, and variance \noisesigma{}.
In most typical settings, $\noise{}(t)$ will be dominated by photon shot noise, making \noisesigma{} approximately proportional to $\E\left[\obs{}(t)\right]$. More detailed analyses of noise in this model can be found in our appendix and supplemental material.}

% \aadd{Here, $\noise{}(t)$ represents any deviation from the expected value of $\obs{}(t)$, which may encapsulate multiple sources of noise. We note that Poisson-distributed components of $\obs{}(t)$ are easily represented as the sum of a mean that contributes to $\light{}\reflectance{}$ and variations that contribute to $\noise{}(t)$. Our supplemental material includes an in-depth analysis of Poisson-distributed photon shot noise for more detail.}

% \aadd{Our plausible manifold can then be described with a distribution over frame values, where each frame has a mean $\light{}\reflectance$ and variance \noisesigma{}. For most of the working range of the image sensor, $\noise{}(t)$ is dominated by photon shot noise, whose variance is proportional to the mean brightness $\light{}\reflectance$. We discuss this in much greater detail in Section \ref{sec:uncoded}, the appendix, and the supplemental.}

\subsubsection{Warping the Plausible Manifold}
Now consider what happens when we modulate $\light$ such that it fluctuates by a zero-mean code $\code{}(t)$ drawn from a distribution that resembles $\noise{}(t)$, with variance $\sigma_c^2$. Our video model at time $t$ is now: 
\begin{align}
\begin{split}
\obs(t)&=\coverbrace{\left(\light{}+\code(t)\right)}{illumination}\reflectance{}\;+\;\noise{}(t)\\
&=\cunderbracelines{\light{}\reflectance{}}{uncoded}{light}+\cunderbracelines{\code(t)\reflectance{}}{coded}{variation}
\;+\;
\noise{}(t)
% \cunderbracelines{\noise{}(t)}{noise}{variation}
\end{split}
\label{eq:OneDNCI}
\end{align}
% Here we have factored the effects of our fluctuation into a coded reflection term $\vcodet{}=\reflectance{}\cdot\code(t)$
The distribution for each frame is now different depending on whether we know the illumination code. If we condition on a known value of $\code(t)$, then the mean of our distribution is shifted by $\code(t)\reflectance{}$, effectively warping the manifold of plausible video. If $\code(t)$ remains unknown, then our coded variation resembles additional noise, increasing the variance of each frame as shown in \Fig{} ~\ref{fig:OneDNCI}. This difference represents our information asymmetry.

\begin{figure*}
    \centering
    \includegraphics[width=\textwidth]{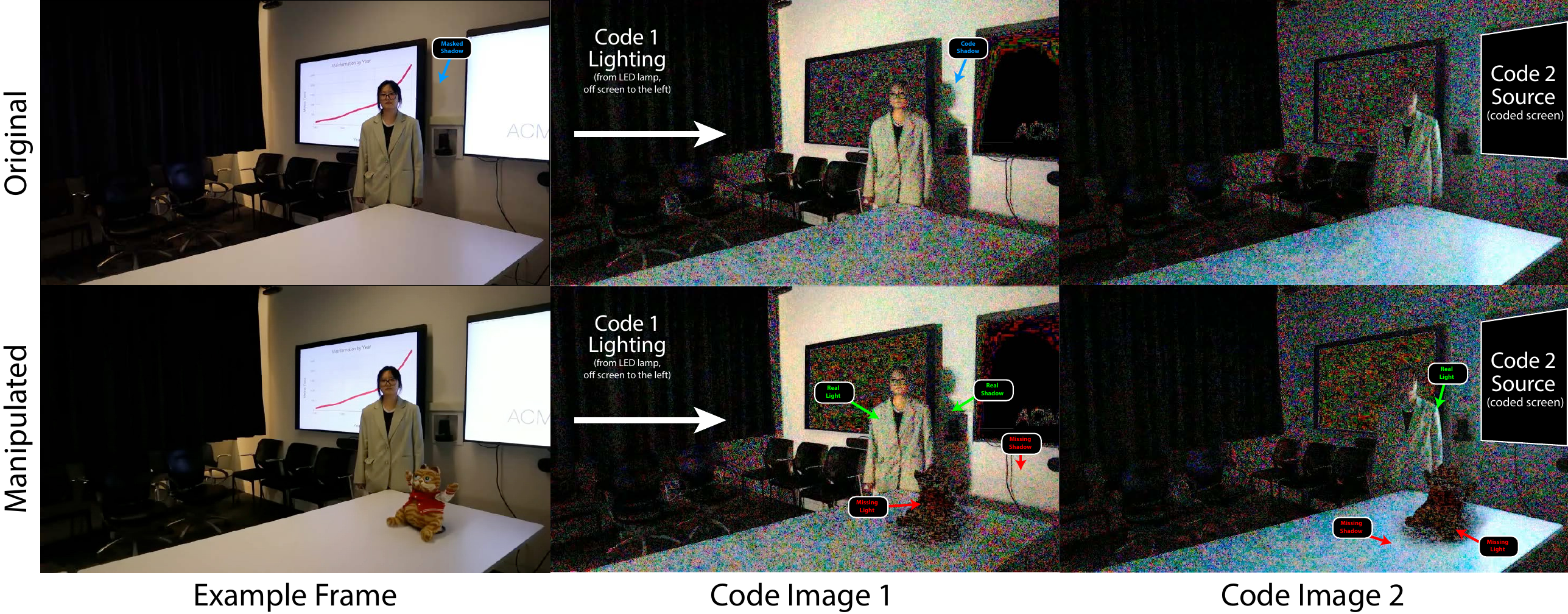}
    \caption{\textbf{Code Images for Lighting Analysis} We can split our code signal bandwidth across multiple light sources to encode a decomposition of their illumination into the apparent noise of a video. In this example, we captured video in a conference room with two coded sources. The first coded source is a LED lamp located off-screen to the left of the scene, and the second coded source is a large monitor partially visible on the right side of the scene. 
    The top row visualizes images extracted from the original video, and the bottom row visualizes corresponding images from a video that has been edited to place a stuffed animal on the table. The left column compares frames taken from the two videos. The middle column compares code images for the first code, which should show the scene illuminated only by the LED lamp from the left. The right column compares code images for the second code, which should show the scene illuminated only by the monitor on the right. 
    In code images from the fake video (bottom) we see several signs of manipulation. The added content does not appear to reflect coded light. Code images also make it easier to see shadows that are masked by other light sources in the original frame. The shadow that the person's head casts from the LED lamp is barely visible in the original video due to an uncoded ceiling light. However, in code image 1, all other sources are removed, leaving this part of the shadow clearly visible. The similar code shadows for added content are missing from manipulated code images.
    % Evidence of manipulation can be found in both of the code images in the bottom row. 
    We recommend zooming in to better view these results.}
    \Description{Left to Right (top: original video, bottom: manipulated video): frame from video, code image 1 from left lamp, code image 2 from right monitor.}
    \label{fig:lightingdecomposition}
\end{figure*}

% \begin{figure*}
%     \centering
% %    \includegraphics[width=\textwidth]{fig/experiments/CarolineDecompWManipPageWidth.pdf}
%     \includegraphics[width=\textwidth]{fig/compressed/CarolineDecompWManipPageWidthAug9_138Comp.pdf}
%     \caption{\textbf{Code Images for Lighting Analysis} We can split our code signal bandwidth across multiple light sources to encode their illumination decomposition into the apparent noise of a video. In this example, we captured a video in a conference room with two coded sources. The code images in the middle correspond to a coded lamp placed offscreen to the left. The rightmost code images correspond to the coded monitor partially seen to the right of the frame. Notice how the inserted plushie results in inconsistencies in the code images although it looks plausible in the video frame. Note that the left television is uncoded, hence it is dark in the code image, which an analyst should keep in mind.}
%     \label{fig:lightingdecomposition}
% \end{figure*}

\subsection{Decoding Illumination}
\label{sec:decodingillumination}
Much of our analysis will concern inner products 
between our observation over some time window $\window{}$ 
and different segments of our known code signal.
% This technically introduces a small correlation between  As this expectation is dominated by uncoded light, our derivation first approximates . We discuss the implications of this in much greater detail in our appendix and supplemental material, and examine its empirical impact in Section \ref{sec:uncoded}. To simplify the derivation of our basic approach, we treat \noise{}(t) as uncorrelated with mean intensity, which is  strong  with our signal, which is.
For notation, we will use bold versions of a variable to denote a vector of its values over $\window{}$  (e.g., $\code\rightarrow\codew{}$, $\light\rightarrow\lightw{}$),
% $\lightw{}$ to denote the value of $\light{}$ broadcast as a vector over our observed time window, and 
and $\coded{}$ will denote a separately chosen segment of our code being used for analysis (potentially different from the observed segment $\codew{}$).
% Finally, we will use $\codedn{}$ to denote the unit vector obtained by normalizing vector $\coded{}$, so that $\codedn{}^\intercal\mathbf{v}$ gives the projection of $\mathbf{v}$ onto $\coded{}$. 
% Projecting our windowed observations $\obsw{}$ onto a chosen analysis segment $\codedn{}$, we get:
Taking the inner product of our windowed observation $\obsw{}$ and one such analysis segment gives:

\begin{align}
\codedd{}^\intercal \obsw{} 
&=
\codedd{}^\intercal
(
\coverbracelines{
\lightw{}\reflectance{}
\;+\;
\noisew{}
}{uncorrelated}{with code}
\;+\;
\codew{}\reflectance{}
)
\\
% (\coded{}^\intercal\lightw{})\reflectance{}
% \;+\;
% % \left(\coded{}^\intercal\codew{}\right)\reflectance{}
% \alignmentTerm{}
% \;+\;
% \coded{}^\intercal\noisew{}\\
&= \cunderbrace{\codedd^\intercal[\lightw{}\reflectance{}+\noisew{}]}{zero-mean}+\cunderbracelines{\alignmentTerm}{alignment}{term}
\label{eq:decode}
\end{align}

\noindent We call $\alignmentTerm{}$ our alignment term because the expectation $\E\left[\codedd{}^\intercal \obsw{}\right]$ over our noise $\noisew{}$ is significant only when the analysis segment is aligned with our observed segment of code:
\beq
\E\left[
\codedd{}^\intercal \obsw{}
\right] =
    \begin{cases}
    \alignedAlignmentTerm{}, & \text{when} \; \coded{} = \codew{}\\
    \approx 0, & \text{otherwise}
    \end{cases}
\label{eq:alignment}
\eeq
\noindent This holds so long as our code is sampled i.i.d. from a zero mean distribution. Our strategy for detecting temporal manipulation will amount to maximizing the alignment term over different choices of $\coded{}$ (Section ~\ref{sec:temp}). Our strategy for detecting spatial manipulation will leverage the ability to decode the transport coefficient $\reflectance{}$ when $\coded{}=\codew{}$. Dividing both sides of \Eq{}~\eqref{eq:alignment} by $\coded{}^\intercal\codew{}$ when $\coded{} = \codew{}$ gives us:
\beq
\E\left[
\frac{\codewd{}^\intercal \obsw{}}{\codewd{}^\intercal\codew{}}
\right]=\reflectance{}
\label{eq:singlepixcodeimage}
\eeq
which depends only on the coded light in our scene.

\subsection{Multiple Light Sources}
% Our derivation so far has made no assumptions about where light originates, only that our observations can be factored into three different types of light, $\{\LConst{}, \CodedSum{}, \NoiseSum{}\}$, where $\LConst{}$ is constant, $\CodedSum{}$ describes variation that is correlated with our code, and $\NoiseSum{}$ describes variation that is orthogonal to our code. Each of these pieces may combine light from different physical sources and paths without affecting our analysis. 

% Our derivation so far has made no assumptions about where light originates, only that our observations can be factored into a constant portion $\LConst{}$, temporal variation that consists of a coded component $\code(t)\reflectance{}$ and an uncorrelated component $\noise{}(t)$. Each of these pieces may combine light from different physical sources and paths without affecting our analysis. 
% \Eq{}~\ref{eq:alignment}-~\ref{eq:singlepixcodeimage}.
% so long as any additional lighting introduced to the scene is uncorrelated with our code. 

% Our derivation so far assumes one light source. We can extend this to an arbitrary amount of uncoded and coded light sources by factoring all uncoded light into a constant term $L$ and assigning the $k$ coded sources in the scene distinct codes $\{\codei{1},\codei{2},...,\codei{k}\}$.

\revcomment{Added motive for multiple sources}
Our derivation above considers a single light source, but
\aadd{we can create even more information asymmetry by applying unique codes to different light sources in the same environment.}
% using multiple coded lights employing multiple unique codes.
% \padd{Using multiple coded light sources, each corresponding to a unique code signal, increases information asymmetry.}
% , but \Eqr{} makes no explicit assumptions about the origin or distribution of light. 
To extend \padd{our formulation} to an arbitrary number of uncoded and coded light sources, we first observe that the addition of uncoded constant light $\lighti{i}$ and photon noise $\noise{}_i$ from a new source $i$ does not change the expectation evaluated in \Eqr{eq:alignment}. 
To simplify our notation, we will use $\LConstw{}$ to describe the sum of all uncoded (i.e., constant) light, and $\noisesum{}$ to describe noise aggregated over all sources. Now we consider the case of $k$ coded light sources with distinct codes $\{\codei{1},\codei{2},...,\codei{k}\}$.
% We can use this observation to assign different codes 
% $\{\codei{1},\codei{2},...,\codei{k}\}$ to different light sources.
% $\{\codeit{0},\codeit{1},...,\codeit{k}\}$, 
Each code $\codei{i}$ will also have a corresponding transfer coefficient $\reflectancei{i}$. If we choose codes that are mutually uncorrelated, then the transfer coefficient of each code can be recovered and analyzed independently. To see this, we rewrite \Eq{}~\eqref{eq:decode} to analyze code $\codei{i}$ in the presence of other codes:
% We will use $\LConst{}$ to represent all constant lighting in our scene, and $\noisew{}$ to represent all noise, again noting that the division of these components across different physical sources does not impact our analysis:
% . When $\codei{j}\perp{}\codei{i}\;\forall{}\;j\neq{}i$, we have:
\begin{align}
\codedid{i}^\intercal \obsw{} 
&=
\codedid{i}^\intercal\LConstw{}
\;+\;
\codedid{i}^\intercal\noisesum{}
\;+\;
\reflectancei{i}\sum_{j=1}^k\cunderbracelines{\codedid{i}^\intercal\codewi{j}}{code}{correlation}
\label{eq:decodemultiplecodedsources}
\end{align}
If we choose uncorrelated codes, 
% (i.e., $\codei{j}\perp{}\codei{i}\;\forall{}\;j\neq{}i$), 
the expected value of our code correlation term $\codedid{i}^\intercal\codewi{j}$ will be zero whenever $i\neq{}j$, and we can perform the analysis of each code independently just as we did with \Eq{}~\eqref{eq:singlepixcodeimage} in our single-code scenario:
\beq
\E\left[
\frac{\codewid{i}^\intercal \obsw{}}{\codewid{i}^\intercal\codewi{i}^{}}
\right]=\reflectancei{i}
\label{eq:codeisinglepixcodeimage}
\eeq
% Another way to interpret this is that adding a second uncorrelated code has the same impact on our expectation analysis as adding more noise.

% All terms involving codes other than $\codei{i}$ are zero mean, leaving 
\noindent The use of uncorrelated codes in this manner is similar to frequency division multiplexing in wireless communication, which divides transmission bandwidth across multiple channels \cite{freqmultiplex}. In our case, each channel carries the transfer coefficient $\reflectancei{i}$ for light modulated by a different code.

% If we have $\nlights{}$ coded sources, we can split our carrier bandwidth across $\nlights{}$ corresponding code signals $\{\codei{1},\codei{2},...,\codei{\nlights{}}\}$ to encode a full decomposition of these sources in the apparent noise of our video as shown in \Fig{} ~\ref{fig:lightingdecomposition}). This results in a significant reduction of the plausible manifold, even in a pathological scenario where the illumination codes are given to an adversary, as the adversary would still have to fake $\nlights$ light sources independently, which is strictly harder than the usual task of faking their sum. 

% \begin{figure}
%     \centering
%     \includegraphics[width=\linewidth]{fig/experiments/CarolineDecomp.pdf}
%     \caption{\textbf{Lighting Decomposition} We can split our code bandwidth across multiple light sources to encode a decomposition of the illumination in a scene into our noise. Here, we captured a video (above) in a conference room with two coded sources. The bottom left code image corresponds to a coded lamp placed to the left of the room. The bottom right shows a code image corresponding to the monitor partially seen to the right of the frame.}
%     \label{fig:lightingdecomposition}
% \end{figure}

\subsection{Code Images}
Our analysis so far includes no limiting assumptions about spatial or color resolution, which makes extending it to these dimensions fairly straightforward. Using $\pix{}$ to index over pixels and channels, we get:
\beq
\obsw(\pix)=\LConstw{}(\pix)
\;+\;
\noisesum{}(\pix)
\;+\;
\sum_{i=1}^k\codewi{i}\cdot\transportiat{i}
\eeq
which lets us update Eq. ~\eqref{eq:codeisinglepixcodeimage} to express an image $\transportiat{i}$ of transport coefficients for code $i$:
\beq
\transportiat{i} = \E\left[
% \frac{\codewid{i}^\intercal \obsw(\pix)}{\codewid{i}^\intercal\codewi{i}}
\frac{\codewid{i}^\intercal \obsw(\pix)}{\codewid{i}^{\intercal}\codewi{i}^{}}
\right]
\label{eq:singlesourceimage}
\eeq
We can think of $\codeimagei{i}$ as an image of the scene lit only by illumination coded with code $\codei{i}$.

\subsection{Radiometric Ambiguity}
\label{sec:radiometric}
In theory, the code $\codewi{i}$ should be measured in radiometric units. This information can be obtained by calibrating light sources, but this is not strictly necessary. Using a code for analysis that is only proportional to the actual amplitude will produce code images equal to $\transportiat{i}$ up to a global scale factor, which is already very useful for forensic purposes and makes \nci{} easier to deploy with minimal modifications to common hardware. The main caveat is that calibration can provide extra protection against attacks from an informed adversary, as we discuss in the supplemental.
% setups to      and since we generally do not know the precise exposure parameters of video being analyzed already, one additional source of global ambiguity makes little difference.
% $\codedi{i}=\codewi{i}/||\codewi{i}||$ in our analysis, resulting in one global scale ambiguity.
% However, as we later show, what matters in analysis are the relative proportions of light transport, so a global scale ambiguity is typically not a problem.
% The resulting code images are then analogous to regular images taken with a fixed but unknown exposure, with each pixel constraining a product of illumination and reflectance.
% Eq. ~\ref{eq:singlesourceimage}
% as a way to calculate an image of the scene where all uncoded light has been removed.
% \coded^\intercal[\lightv{}\transportat{\pix}+\noisew{}]+\left(\coded{}^\intercal\codew{}\right)\transportat{\pix}

% We can use code images to detect spatial tampering by looking at how the scene looks under the sole illumination of each coded source, and look for inconsistencies like in Figure~\ref{fig:teaser}.

\section{Analyzing Video}
\label{sec:vidanalysis}
\subsection{Global Temporal Registration}
\label{sec:alignment}
The first step in our analysis is to align a video of interest with the code signal. We start by calculating a global vector $\obsaw{}$, which averages our observations $\obsw(\pix)$ over all pixels and channels $\pix$.
% of observed values averaged over the pixels and channels in a $t$.
% $\propto\sum_{\pix}\obsw{}(\pix)$
% to refer to a time-sequence of observations averaged over the pixels and channels of each frame. a time series vector of pixel averages calculated for each frame in our video.
% Our observed segment of the code, $\codewi{i}$, is calculated as the segment that maximizes the corresponding alignment term derived from Eq.~\eqref{eq:alignment}:
We find the segment of code $\codewi{i}$ corresponding to our observation by maximizing the corresponding alignment term from Eq.~\eqref{eq:alignment}:
\beq
\codewi{i}=\argmax_{\codedi{i}}\;\codedi{i}^\intercal\obsaw{}
\eeq
This can be solved with simple matched filtering: we perform cross-correlation between our code signal and $\obsaw{}$, then pick the segment of our code corresponding to the peak correlation. This method works even in the presence of significant spatial manipulation (e.g., \FigRef{fig:teaser}), extremely low code signal levels (e.g., amplitudes averaging less than one brightness level per pixel in video), and in video with very high compression rates (see supplemental for detailed experiments and analysis). Thus, failure \padd{of global temporal registration} in a video of a scene where most light sources are coded is a strong indicator of either temporal manipulation or frame-wide spatial manipulation.

% average valan average signal $\obsg{}(t)$
% We do this by performing matched filtering between the code signal and the time series obtained by spatially averaging each frame in the video.

% \subsection{Characterizing Temporal Manipulation}
% The first step in our analysis is to identify which segment of code is present in a video. To do this, we that is present in a video.
% Before we can compute code images, we need to align a captured video with our known illumination code. To do this, we perform matched filtering between the code signal and the time series obtained by spatially averaging each frame in the video. If the video does contain the signal, a sharp peak will appear in the result of the matched filter with its corresponding time delay indicating the phase of the coded signal in the video as shown in Figure \ref{fig:matched_og}.
% Figure \ref{fig:matched_og}

\subsection{Characterizing Temporal Manipulation}
\label{sec:temp}
When global registration fails due to temporal manipulation of a video, we can use \nci{} to further characterize several common types of tampering, including:
\begin{itemize}
\item \textbf{Malicious Cuts:} Arguably the most common way to spread misleading information through video is by simply editing together segments that leave out or distort context. This can be done by jumping forward or backward in time when cutting normally between shots (e.g., to change responses in interview footage) or by creating "warp cuts", which synthesize interpolating frames in order to hide a cut.
Recent examples of disinformation spread with malicious cuts include viral videos of Joe Biden \cite{bidenphysical} and former Senator Dianne Feinstein \cite{feinstein} prior to the 2020 election.
% prior to the 2020 election  and one of former Senator Dianne Feinstein Recent examples of disinformation spread with malicious cuts include a video of then Presidential-Candidate Joe Biden % seeming to call for a physical revolution in 2019 \cite{bidenphysical} and one of
% seeming rude and dismissive to a group of schoolchildren when talking about climate change .

\item \textbf{Speed \& Acceleration Changes:} Changing the speed of a video can be used to change an individual's apparent behavior. One famous example of this was a viral fake video appearing to show Representative Nancy Pelosi intoxicated \cite{PelosiSlur}. Localized speed changes can be used to make an individual seem more aggressive, like when part of a video was sped up to make it seem like journalist Jim Acosta struck a White House staff member during a press conference \cite{AcostaFake}, which was used by the Trump administration as justification for suspending his press credentials.
\end{itemize}

\noindent We can identify each of these manipulations by searching over a more general space of alignments with our video.

\subsubsection{\AMatrices{}}
% We can identify localized temporal manipulations done to video, such as malicious cuts and local changes in speed, by searching for local alignments of a video with our code.
% If temporal manipulation causes the alignment with our code signal to fail, we can characterize
We can characterize localized temporal manipulation by searching for local alignments of a video with our code.
% in which case we search for local evidence of our signal, which serves as a way to identify the nature of temporal manipulation as well.
% We do this by calculating cross-correlation with a sliding window of video segments, which
% divide the video into small overlapping windows and use cross-correlation to find the alignment of each window. This
We do this by calculating a matrix of alignment scores, where each column is a cross-correlation between the corresponding segment of video and our code,  as shown on the right of \Fig{}~\ref{fig:teaser}. This representation is similar to the correspondence matrix used in dynamic time warping (DTW). Unmanipulated segments of a video should show up in an \amatrix{} as strong unit-slope diagonal segments representing the most likely alignment of each frame with its corresponding time in our code. We call the set of such segments in a matrix our \emph{\alignmentcurve{}}, which indicates the map from each video frame to the corresponding time that it was captured. Any localized temporal manipulation will create a discontinuity in the \alignmentcurve{}. When a cut jumps forward or backward in time, we can tell how much time was removed or added by looking at the offset between discontinuous segments. Temporal manipulation is usually easy to see in \alignmentcurve{}s, and our supplemental describes a metric for optionally automating its detection.
% Searching for discontinuities in \alignmentcurve{}s forms the basis of our automatic temporal tampering detection algorithm proposed in the supplemental, whose results are immediately above most of the correlation matrices shown in this paper.
% some variation from this diagonal.
% , which is a basis for our automatic detection algorithm proposed in Section \ref{sec:auto}. The characteristics of a discontinuity can also give us clues about the type of tampering done.
\Fig{}~\ref{fig:interview} shows the alignment matrix for a video that has been maliciously cut to rearrange segments in time, with the specific rearrangement clearly visible in the matrix.
% (e.g., \Fig{}~\ref{fig:teaser}, bottom right \abe{change this to new figure and describe}).
% offset between  while

% a more robust strategy for characterizing speed changes is to first identify regions where they may occur, and then it is more robust to first identify  strategy for characterizing these changes is to  can perform a subsequent analysis on the gap region using a smaller time window to further characterize local speed changes.
% However, this comes with at a cost to SNR, so it is safer to first identify gaps using a larger window, then perform additional local analysis around individual gaps.
% , but  in the form of curves that bridge these low energy regions. However, it is curves that connect in the alignment matrix) at the cost of SNR, but we instead .
% a sudden change in the diagonal as \Fig{} \ref{fig:teaser} shows. Local changes in speed will result in a similar artifact following an ambiguous, low-energy region (since cross-correlation is not equivariant to timescale).
% or global speed changes will resultsresult in segments with lower correlation, and cuts as linear cross-correlation is only robust to minor temporal stretcannot align signals of different timescales. Additionally, the correlations before and after the tamper will not be co-linear in this case as well as with temporal cuts.

\subsubsection{Characterizing Speed Manipulation}
\label{sec:characterizingspeedmanipulation}
Small changes in the speed of a video ($<~5\%$) can often be detected as changes in the slope of an \alignmentcurve{}, but cross-correlation has limited robustness to time scaling by larger factors.
% , so changes to global video speed will cause the entire \alignmentcurve{} of a correlation matrix to lose energy. 
We can characterize a wider range of speed changes more robustly using a correspondence search over time scales.
% Changes in speed can often cause cross-correlation to fail, but can be detected with a search over different time scales.
% , like the ones used to make former Speaker Nancy Pelosi appear intoxicated \cite{PelosiSlur},
% will often fail to register with cross-correlation.
% will often cause cross-correlation-based global alignment to fail, but we can still identify these changes with \nci{}.
A na\"ive search of this sort would be very slow, as it covers a 2D space of time scales and alignments. However, we can factor this search into a 1D optimization over timescales in the frequency domain followed by a 1D optimization over alignments in the time domain.
% as previously described in \Sectn{\ref{sec:alignment}}. 
Recall that changing the speed of a signal by some factor corresponds to scaling its frequencies by the inverse of that factor in the Fourier domain.  We first find the speed change $\speedFactor{}$ by maximizing an inner product of magnitudes from $\Fobsaw$, the Fourier transform of our spatially-averaged observation $\obsaw$, and $\FCodew{}_{\speedFactorCandidate{}}$, the Fourier transform of our code signal
% the sum of the code signals $\codeav{}$ 
time-stretched by candidate factor $\speedFactorCandidate{}$:
\beq
\speedFactor{}=\argmax_{\speedFactorCandidate{}}|\Fobsaw|^\intercal|\FCodew{}_{\speedFactorCandidate{}}|
% \speedFactor{}=\argmax_{\speedFactorCandidate{}}\sum_{\freq}|\Fobsa(\freq)|\cdot |\FCode{}(\freq/\speedFactorCandidate{})|
\label{eq:speedfactordetection}
\eeq
% correspondence between the DFT amplitudes for $\obsaw{}$ and stretched versions of $\codeg{}$.
Once we have found the correct speed factor, we can perform matched filtering in the time domain with the corresponding resampling of our code to find the correct phase shift. The top row of \Fig{} \ref{fig:dtw} visualizes the result of our timescale search on video that has been slowed down by a factor of $0.6x$.

\subsection{Characterizing Spatial Manipulation}
Most malicious spatial tampering of video falls somewhere on a spectrum between ``shallow" methods (sometimes called  ``cheap fakes"), which rely on direct image or video compositing, and ``deep" methods (i.e., ``deep fakes"), which rely on generative models trained with significant amounts of data. The majority of past incidents (at the time of writing this paper) fall squarely on the shallow end of this spectrum, but concern about deep fakes is increasing with new developments in generative technology. An illustrative example of a shallow fake is one viral fake video of Joe Biden on a campaign stop from the 2020 presidential election, where a sign was doctored to make it look like he was addressing the wrong crowd \cite{BidenShallowFake}. 
More recently, a deepfake video made headlines appearing to show Ukrainian president Volodymyr Zelenskyy telling his country's armed forces to surrender to Russia \cite{zelensky}. 
The classification of tampering as deep or shallow can be ambiguous, and in either case, our strategy for detecting spatial manipulation is rooted in the analysis of code images. However, as we explore in the supplemental, 
the specific evidence found in code images may differ depending on how the tampering was performed. 

\subsubsection{Detecting Na\"{i}ve Spatial Manipulation}
Most spatial manipulation of video will erase local traces of illumination codes, making manipulated pixels appear black in recovered code images. This is usually easy to detect with the naked eye (e.g. \Figs{}~\ref{fig:teaser} and ~\ref{fig:lightingdecomposition}). However, dark and shadowed objects may also appear black in code images, so it is important to consider dark regions of a code image in context. For this, it often helps to consider the corresponding pixel values in the original video. If pixels appear
well-exposed to a coded source in the original video, but dark in the corresponding code image, this is a strong indication of manipulation. We propose a new visualization based on this idea in the supplemental. 

% \abe{Potentially refer to LC ratio vis}
% For example, if a dark code region is clearly exposed to a coded light source, it is likely fake or manipulated. If the scene may contain a mix of coded and uncoded light, the analyst should reason about how much coded illumination would fall on relevant parts of the scene.

\subsubsection{Forensic Lighting Analysis}
 Lighting and shadows are among the hardest things to fake in video, and have been the focus of forensic analysis strategies in the past (e.g., \cite{faridshading,faridshadows}). 
 With \nci{}, each code image offers a view of the scene under different lighting conditions that can support this type of analysis, providing a very robust forensic advantage. For example, many scenes with multiple light sources contain what we call \emph{\codeshadows{}}, which are shadows cast by a coded light source onto surfaces that appear bright due to illumination from other sources. These \codeshadows{} may be barely visible in the original scene, but appear clearly in code images. The body of the person in our politician scene (\Fig{} \ref{fig:teaser}) and the head of the person in our conference room scene (\Fig{} \ref{fig:lightingdecomposition}) both cast code shadows, but fake content added to these scenes does not, indicating that it was not present in the original video. 
 \aadd{One could conceivably automate the detection of such inconsistencies, but the intepretability of coded shadow artifacts makes it easier to integrate their use as part of a broader forensic strategy.}
 \revcomment{Clarified visual/ interpretable analysis}
 % These shadows are not cast by the fake content added to each scene, which is a clear sign of manipulation in code images.
 
 Notably, even in the extreme case where an adversary knows that \nci{} is used and somehow gains access to the illumination codes, our approach still offers a forensic advantage by way of a manifold reduction. This is because, even if we remove information asymmetry, faking a decomposition of lighting for manipulated content is strictly more difficult than faking a single image.

\section{Designing the Code Signal}
\label{sec:design}
The only assumptions we have placed on our code signals so far are that they should be random, noise-like, zero-mean, and uncorrelated with each other. 
Subject to these restrictions, there are two additional and often competing criteria we would like our codes to satisfy. First, our code should be reliably detectable even under common levels of video compression. And second, to achieve our goal of subtlety from \Sectn{\ref{sec:designgoals}}, coded fluctuations should be difficult for humans to notice. 
% Additionally, while our code should be noise-like, it should be scene and recording device-independent. 
Our strategy for sampling codes will operate in the Fourier domain, where these criteria are simpler to reason about. We will describe how to sample signals in segments, noting that the process can be extended to infinite codes by seeding the generation of each segment by a function of its offset from some reference time.

\subsection{Bandwidth}
To resemble noise, our code should be relatively broadband. This will also offer some robustness to compression.
Assuming common video frame rates of 24-\qty{30}{\hertz} we want to band-limit our code safely under a Nyquist frequency of \qty{12}{\hertz}. Also, at low frequencies, there is some risk of triggering the auto-exposure function of digital cameras. With these factors in mind, we band-limit our code to 2-\qty{9}{\hertz}. We note that strategies based on larger bandwidths could be a direction for future work (e.g., related to \cite{roberts2013undersampled}).

% Low-frequency changes in lighting  we band-limit our code signal to \qty{9}{\hertz}, below the Nyquist frequency of \qty{15}{\hertz} for more robust recovery. At low frequencies, there is some risk of triggering the auto-exposure function of digital cameras, so we also filter frequencies lower than \qty{1}{\hertz}. This leaves us a band of \qty{1}{\hertz}-\qty{9}{\hertz}. We note that strategies based on larger bandwidths could be a direction for future work (e.g., related to \cite{roberts2013undersampled}).

% We want the code signal to have similar characteristics to image noise, which is generally broad-spectrum. However, assuming a common video framerate of about 30Hz, we band-limit our code signal to the Nyquist frequency of 15Hz to minimize aliasing. We note that larger bandwidths could be possible \cite{roberts2013undersampled}, but are left to future work. At low frequencies, there is some risk of triggering the auto-exposure function of digital cameras, so we also filter frequencies lower than $1Hz$. This leaves the code in the $1Hz-15Hz$ frequency band.

% The code should also have similar statistics to noise, which is generally broad-spectrum, so we  in our remaining bandwidth. 
% The code should also have similar statistics to noise, which is generally broad-spectrum, so we  in our remaining bandwidth. 

\subsection{Segment Size \& Multiple Codes}
% Even in the case of non-repeating codes, we generate our signal in segments to more easily divide bandwidth across multiple sources. 
% Our segment size controls the 
The number of samples in each segment determines the resolution at which we can control our chosen frequency band.
To generate $k$ codes for a segment, we assign them to frequencies in a round-robin fashion, ensuring that each frequency belongs to exactly one code. This guarantees that the codes will be mutually uncorrelated. We shuffle the order of assignment for each block of $k$ frequencies to ensure that codes cannot be identified by a regular distribution of component frequencies. 
% It also ensures that there is no way to distinguish between the variation of a single code and a superposition of multiple codes.
Shuffled round-robin assignment also has the property of making the sum of our codes independent of $k$, making it impossible to infer the number of illumination codes from temporal analysis alone.

% sum of our codes will be there is no way to distinguish between the variation of a single code and a superposition of multiple codes.

\subsection{Phases \& Amplitudes}
\label{sec:flicker}
The phase component of the coefficient corresponding to each frequency is chosen uniformly at random. We sample the amplitude of each coefficient based on related psychophysical studies on human flicker sensitivity \cite{kelly1961visual, de1961eye}. Specifically, we sample each amplitude from a uniform distribution with a mean that is inversely proportional to human flicker sensitivity at the corresponding frequency, based on estimates from \cite{flicker}. We include additional flicker sensitivity analysis in our supplemental material.

\section{Dealing with Motion \& Uncoded Light}
\label{sec:mouncoded}
Our analysis thus far has not considered the impacts of scene motion or limited precision. Here we describe steps that can be taken to make our analysis fairly robust to these factors.

\begin{figure}[t]
\centering
\includegraphics[width=\linewidth]{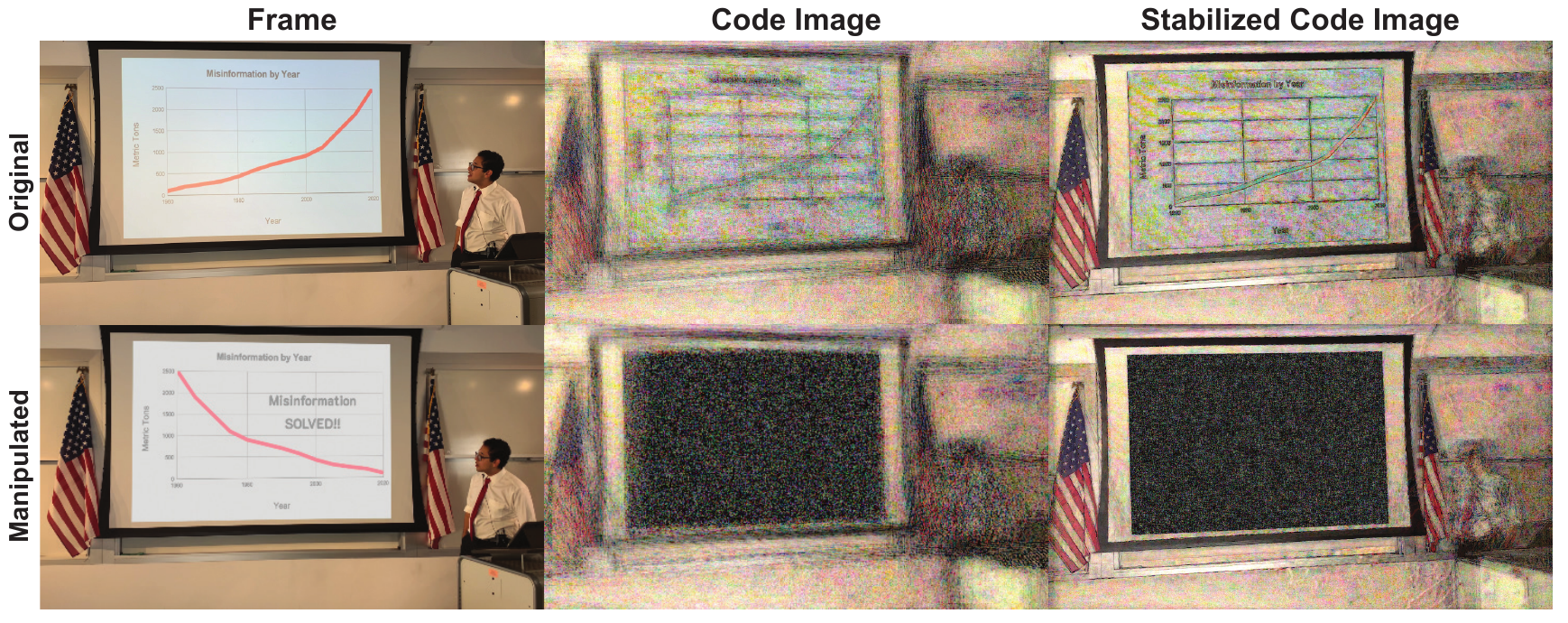}
\caption{\textbf{Stabilization for Hand-Captured Video} Performing standard homography-based video stabilization can greatly improve code images recovered from hand-captured video.
% correct for Code Images from a handheld video recorded on an iPhone XS after stabilization and compression from different popular social media sites. Note the clear anomaly in lighting present in all the manipulated code images. Note that Reddit videos are 30fps instead of 60fps, resulting in half the frames used for computing its code image to maintain the same window duration.}
}
\Description{Left to Right (top: original video, bottom: manipulated video): frame from video, unstabilized code image, stabilized code image}
\label{fig:stabilized_handheld}
\end{figure}

\subsection{Stabilization \& Filtering}
\label{sec:motion}

\begin{figure}[t]
    % \begin{subfigure}{\linewidth}
    %     \includegraphics[width=\linewidth]{fig/bilateral.pdf}
    % \end{subfigure}
    % \begin{subfigure}{\linewidth}
    %     \includegraphics[width=0.49\linewidth]{fig/experiments/bi/pol_no_bi_cimg.png}
    %     \includegraphics[width=0.49\linewidth]{fig/experiments/bi/pol_bi_cimg.png}
    % \end{subfigure}
    \includegraphics[width=\linewidth]{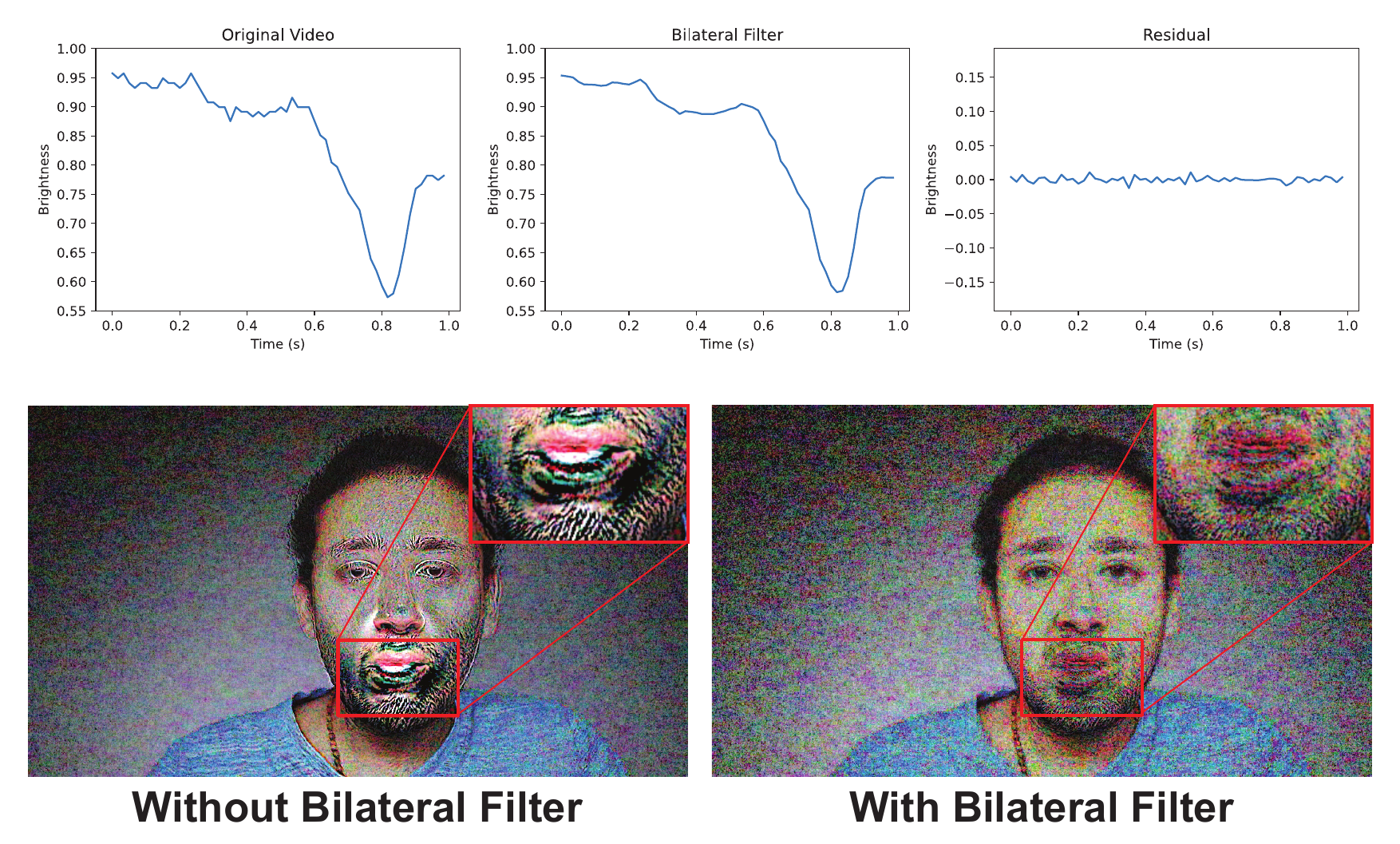}
    \caption{\textbf{Motion Filtering} We use a temporal bilateral filter to reduce the impact of transient changes such as those caused by motion. The top row of plots shows the filter applied to a pixel in the mouth region of a talking human subject. The left plot shows original intensities, the middle shows bilateral filtered values, and the right shows the residual, which we use for our analysis. The bottom row shows code images without (left) and with (right) bilateral filtering applied. Note how the bilateral filter removes the transient response from the subject's mouth motion.}
\Description{Top Left: original pixel values over time. Top Middle: bilateral filtered pixel values over time. Top Right: residual pixel values over time. Bottom Left: code image without bilateral filtering applied. Bottom Right: code image with bilateral filtering applied.}
\label{fig:bilateral}
\end{figure}

% \begin{figure}[ht]
%     \begin{subfigure}{\linewidth}
%         \includegraphics[width=\linewidth]{fig/bilateral.pdf}
%     \end{subfigure}
%     \begin{subfigure}{\linewidth}
%         \includegraphics[width=0.49\linewidth]{fig/experiments/bi/pol_no_bi_cimg.png}
%         \includegraphics[width=0.49\linewidth]{fig/experiments/bi/pol_bi_cimg.png}
%     \end{subfigure}
%     \caption{\textbf{Bilateral Filter for Motion Filtering} We use a bilateral filter for filtering out transients that occur from motion in a video. The top figure shows a real example of a pixel located on the lip region of someone talking. The leftmost plot is the original signal over time. Notice the large jumps in brightness. The middle shows the result after the bilateral filter is applied, yielding primarily the transient we aim to get rid of. Computing the residual, we are left with our code signal. The bottom figure shows code images before (left) and after (right) bilateral filtering is applied. Note how the bilateral filter removes the transient response from the subject's hand motion.}
% \label{fig:bilateral}
% \end{figure}
We have two strategies for dealing with motion. The first way is to align our reference frame with motion of the scene (i.e., use a Lagrangian frame of reference), which we do by stabilizing video prior to analysis. For hand-captured video, we perform homography-based stabilization on the entire frame (\Fig{}~\ref{fig:stabilized_handheld}). For close-up videos of a face, we can also stabilize based on face tracking to help detect deep fakes (see supplemental). Our second strategy for dealing with motion involves filtering the intensity of each pixel over time (i.e., filtering an Eulerian reference frame). Assuming that large changes in the intensity of a pixel signify scene motion, we replace each observed intensity $\obsw(\pix)$ with $\obsw(\pix)-\bilateralof{\obsw(\pix)}$, where $\bilateral{}$ is a 1D temporal bilateral filter. \Fig{}~\ref{fig:bilateral} shows an example of the impact this filtering can have on code images. 
% In this case, filtering removes artifacts around the mouth of a subject caused by  scene with motion around the mouth of a speaking subject. mouth motion from a subject speaking. 
Please refer to the supplemental for implementation details and more ablations, including an example where bilateral filtering removes the impact of camera flash in a captured video.
% We provide more details on these optional steps as well as an ablation in the supplementary material.
% for each stabilized pixel  perform a 1D temporal bilateral filter on each pixel and subtract the resulting values from our loc from   over time for each pixel and subtract the resulting value from

% The Eulerian approach is to root our analysis in  which roughly correspond to Eulerian and Lagrangian reference frames. The  and  Lagrangian and One is to anchor our frame of reference in the scene, so that each local time signal describes let our frame of reference move with
% Our analysis thus far has assumed a static scene.
% Due to the temporal coding used by \nci{}, our analysis assumes a static scene and camera, which is impractical. We discuss some approaches to relax these assumptions.

% \subsubsection{Filtering Motion \& Other Transients}
% Motion, and other factors such as camera flash, can cause rapid changes in brightness that drown out our code signal. We can think of these transients as an edge in an image, but happening in time instead of space. Bilateral filters have been used for decades on images as edge-preserving filters. Applying it temporally preserves the transients we aim to remove (the "edge") while averaging out the code signal (the "noise"). This is the opposite of what we want, so we take the residual of our observation and the filtered result to remove transients and preserve the code signal. We provide more details as well as an ablation in the supplementary material.

\subsubsection{\TransientFiltered{} Code Images}
\label{sec:transient}
\begin{figure}
    \centering
    \includegraphics[width=\linewidth]{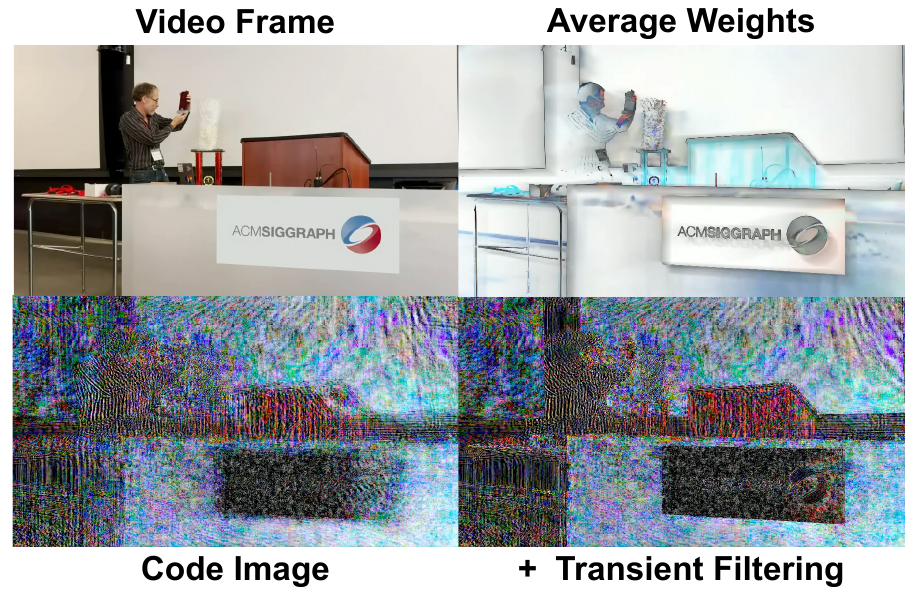}
    \caption{\textbf{Transient-Weighted Code Images} Transient-weighted code images weigh images in the temporal window based on the difference in color from a reference frame in that window when computing code images. This helps maintain the structural similarity between the code image and its corresponding reference frame as shown in the above manipulated example where the SIGGRAPH sticker was added to the desk. The top right image represents the spatially varying mean transient weights. Darker means a smaller effective window size, leading to a lower SNR estimate.}
    \Description{Top Left: manipulated video frame. Top Right: spatially varying mean transient weights. Bottom Left: code image without transient weighting. Bottom Right: code image with transient weighting.}
    \label{fig:motionweight}
\end{figure}

Filtering each pixel with a bilateral effectively prevents transient content (e.g., motion, camera flashes) from being misinterpreted as part of the code signal. However, motion can still cause $\transportiat{i}$ to take on multiple values within our temporal analysis window, causing an artifact analogous to motion blur. To reduce this blur, we can calculate a \emph{\transientfiltered{} code image}, $\motiontransportiat{i}$, based on a temporal window centered at time $t$. We calculate $\motiontransportiat{i}$ with a weighted variant of \Eq{}~\eqref{eq:singlesourceimage}, where the weight $\motionweight(\pix,t,\toffset{})$ of pixel $\pix{}$ at time $t+\toffset{}$ has a Gaussian falloff with its difference in value from the corresponding pixel at time $t$:
% $\motionweightat{x}$ to favor frame values closer to the one present at the center time $t_c$ of our window:

\begin{align}
\motiontransportiat{i} &= \E\left[
\frac{\left(\motionweightatxt{} \odot \codewi{i}\right)^\intercal \obsw(\pix)}{\left(\motionweightatxt{} \odot \codewi{i}\right)^\intercal\codewi{i}}
\right]\\
\motionweightatxt{} &= \left[\motionweightt(\pix{},t,-\frac{\window}{2}),...,\motionweightt(\pix{},t,0),...,\motionweight(\pix{},t,\frac{\window}{2})\right]
\\
\motionweight(\pix,t,\toffset{}) &= \exp\left[ -\frac{1}{2} \left(\frac{\obs(\pix,t+\toffset{}) - \obs(\pix,t)}{\sigma}\right)^2 \right]
% \N(\;\obs(x,t+\toffset);\; \obs(x, t), \sigma)
\label{eq:tfcodeimage}
\end{align}

\noindent Here $\odot$ is an elementwise product, and the weights of any saturated pixels are also set to zero. Smaller $\sigma$ values provide stronger transient filtering at a potential cost to SNR. We can think of this as a spatially varying adaptive window size that filters out pixels with variation significantly above our expected code amplitude. We use $\sigma = 0.05$ (on a $[0,1]$ scale) for most examples. 
The mean weight used for each pixel (visualized in the top right of \Fig{}~\ref{fig:motionweight}) is analogous to the size of our adaptive window and can be used to assess the reliability of pixels
% Visualizing the mean weight used for each pixel (\Fig{}~\ref{fig:motionweight}, top right) gives us a reliability measure 
when assessing code images. The videos in the supplemental website better demonstrate the effectiveness of \transientfiltered{} code images.

\subsection{High Levels of Uncoded Light}
\label{sec:uncoded}

\begin{figure}
    \centering
    \includegraphics[width=0.386\linewidth]{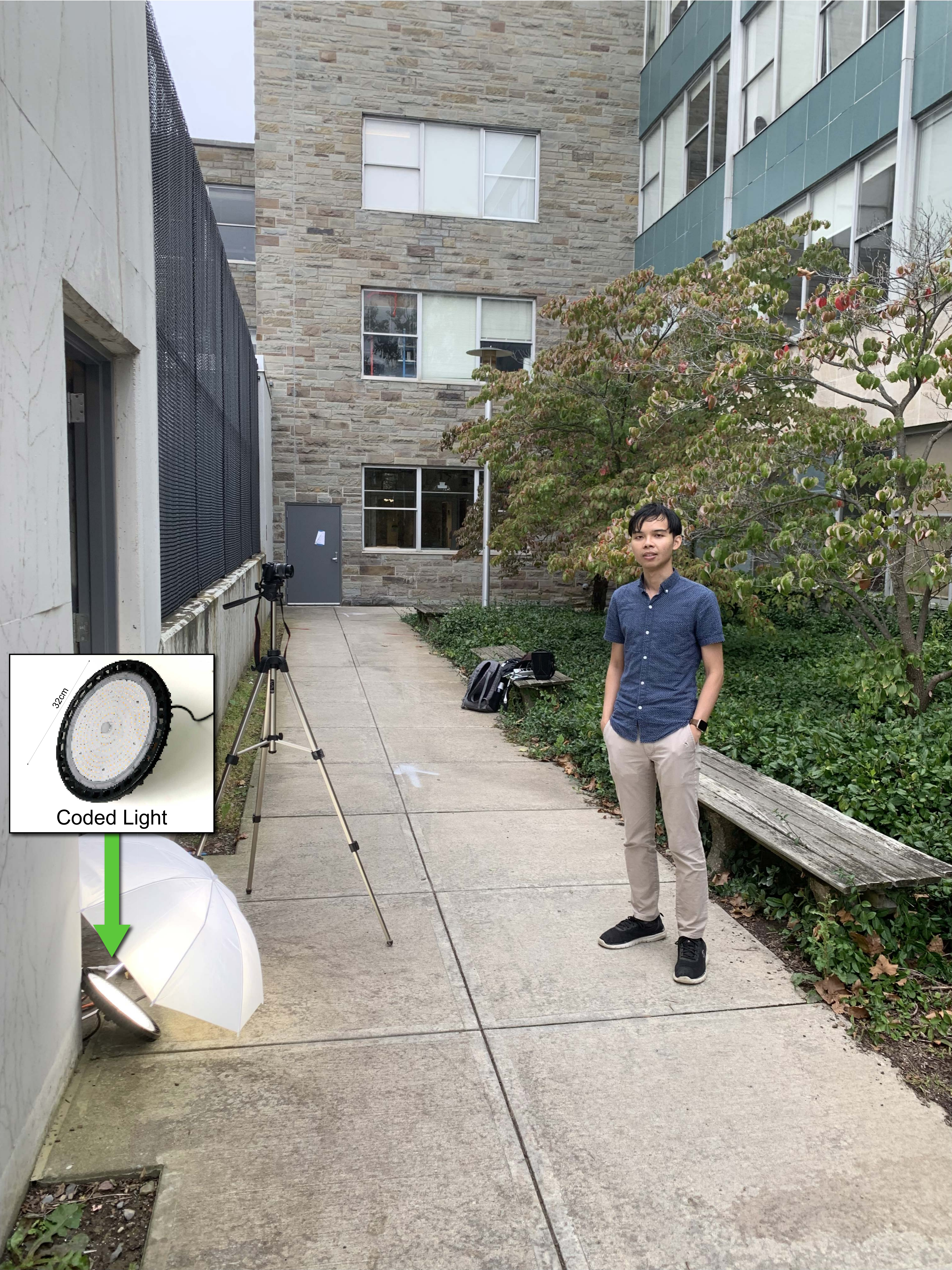}
    \includegraphics[width=0.29\linewidth]{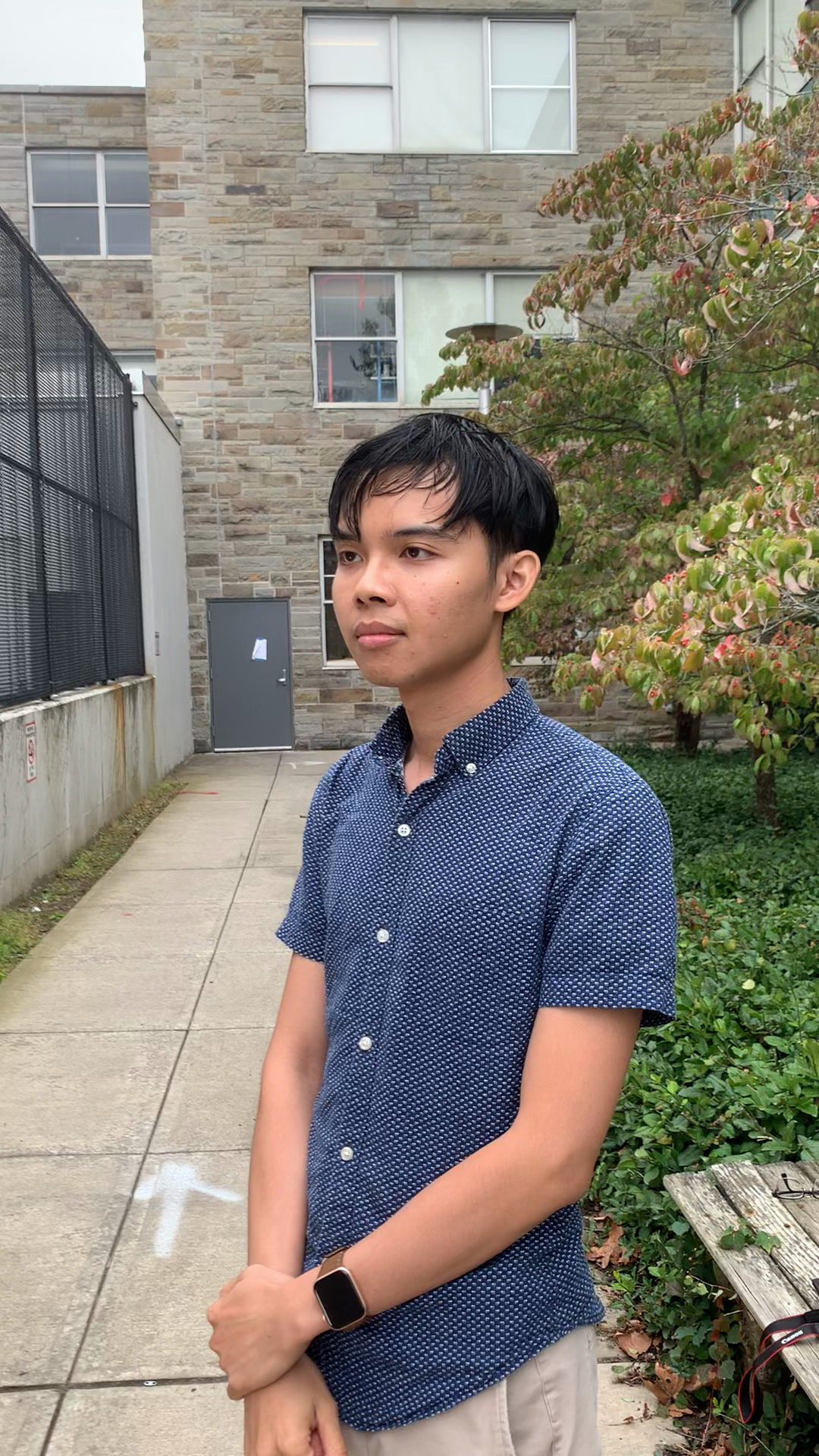}
    \includegraphics[width=0.29\linewidth]{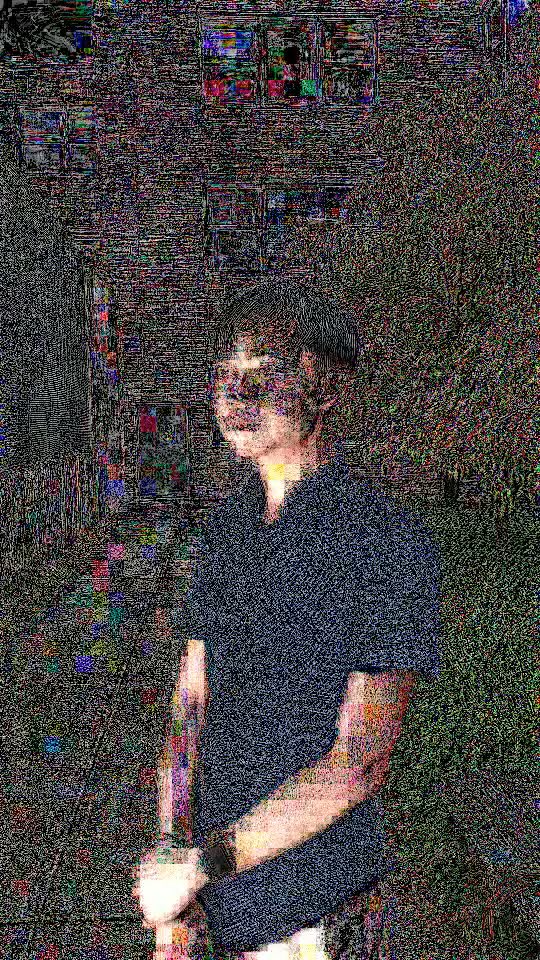}
    \Description{Left: outdoor capture setup. Middle: example video frame from the outdoor capture. Right: code image from the outdoor capture.}
    \caption{\textbf{Outdoor Capture Setup} The left image shows our setup for outdoor capture. The coded light source is on the bottom left under a white umbrella diffuser. Note that the mounted camera was not used for this experiment. Instead, we used a handheld iPhone XS. An example video frame and code image from this setup are shown in the middle and right, respectively.}
    \label{fig:outdoor_setup}
\end{figure}

Our analysis so far has leveraged the assumption that our code is uncorrelated with the noise $\noisew{}$ in our scene (\Eq{}~\eqref{eq:alignment}). 
% More specifically, \Eq{}~\eqref{eq:alignment} assumes that the expectation of $\codedd^\intercal\noisew{}$ is zero. 
However, the SNR of our estimate of $\reflectancev{}$ depends on the variance of $\codedd^\intercal\noisew{}$, which scales with the variance of $\noisew{}$. The primary contributor to $\noisew{}$ is photon shot noise, which is Poisson distributed, meaning the variance of $\noisew{}$ scales approximately linearly with the amount of uncoded light in a scene. Our supplemental material and appendix include a detailed analysis of photon shot noise in \nci{}. In practice, photon noise should be considered more carefully in scenes with high amounts of uncoded light---especially outdoor scenes.
In sunlit settings, the average noise level of pixels may be higher than the amplitude of our code. However, in such scenarios, the local SNR of our code signal spikes in shaded regions and around surfaces that face away from  the Sun (see \Fig{}~\ref{fig:outdoor_setup}). In this case, we run a modified version of our initial temporal alignment that divides the video into patches that are then weighted according to local SNR. We detail this approach in the supplemental, and find that it greatly improves recovery in challenging outdoor conditions.

We can observe a more minor, localized effect of photon noise in the reflection of uncoded ceiling lights visible in (\Fig{} \ref{fig:teaser}). These reflections appear dark and noisy in recovered code images, due to a combination of clipping and high photon noise.

% In such cases, the local SNR of our code signal can vary drastically across different pixel regions, depending on the relative amount of sunlight and coded illumination. we are still able to recover practical signal levels from shaded regions, assuming the coded light source is not coming from the same angle as sunlight (see \Fig{}~\ref{fig:outdoor_setup}). Even in these challenging settings, we can still use many features of \nci{} by splitting a video into local patches and focusing analysis on patches with higher SNR. We detail this approach in the supplemental, and find that it greatly improves recovery in challenging outdoor conditions.

\section{Implementation}
\label{sec:method}
Our analysis code is implemented in Python and PyTorch. 
Our microcontroller code for controlling light sources is written in C for low latency.
% for GPU processing (if available). We will release our codebase upon publication.

\subsection{Hardware}
\label{sec:hardware}
\nci{} can be used with fairly minimal adjustments to common inexpensive hardware. We experimented with two implementations. Our first approach is to fluctuate the brightness of standard computer or television screens, optionally using different codes for different portions of the screen, to create multiple sources. This approach requires no hardware modifications and can be implemented on top of existing screen content.
% Thanks to the relatively low-frequency content of the code signal, \nci{} does not require elaborate hardware design. We have achieved good results with two different low-cost designs. The first fluctuates the brightness of pixels on a standard 60Hz computer screen, which we can divide into multiple sections to create multiple coded light sources. This could easily be deployed as a downloadable application that blends with existing screen content, such as a teleconferencing call. 
Our second approach uses a popular, low-cost ($<50$ USD) \MYhref{https://www.amazon.com/dp/B0BWY42F1P?th=1}{commercially available LED light} with a peak brightness of 28,000 lumens. Most modern light fixtures adopt a ``0-10V dimming'' standard~\cite{ansi2022dimming}, where the brightness of the light is controlled by the potential difference across a pair of wires. Directly injecting the code signal into this dimming system does not work, as the light's internal circuitry contains a low-pass filter with a sub-\qty{1}{\hertz} cutoff. We fix this by adjusting the filter components and bypassing its internal pulse-width-modulation (PWM) section, a simple modification for manufacturers to incorporate. This gives a bandwidth of over \qty{100}{\hertz}, which is more than enough for our code signal. We use an ESP-32 microcontroller \cite{esp32} running our compiled C code to modulate the light with our code signal, which gives us interactive control over brightness and signal amplitude.

\section{Experiments \& Results}
\label{sec:experiments}
We perform qualitative and quantitative analysis on an extensive set of real and synthetic experiments. Many of our experiments are included in the supplemental material.
% We focus primarily on qualitative assessments of real-world experiments in the main paper, and refer to the supplemental for more detailed quantitative analyses.

\subsection{Simulations}
\label{sec:synthetic}
As \nci{} is a new approach, our quantitative analysis focuses on understanding how different factors impact the SNR of recovered codes. 
% The bulk of this analysis can be found in our appendix and supplemental material.  
% and we focus on qualitative assessment of experiments that resembles realistic use cases.
% \nci{} represents a new imaging modality for video forensics. To better understand this modality, 
In addition to our real experiments, we use simulations based on ray tracing in Blender to examine several of these factors under precisely controled conditions. These experiments and their results are detailed in our supplementary material, including experiments that explore the impact of:
% Some aspects of \nci{} are easier to study in simulation, where various properties of the scene can be controlled exactly. Our supplemental material includes synthetic experiments examining robustness to:
\begin{itemize}
    \bulletitem{Code Signal Strength \& Flicker Perception:} We examine the relationship between code signal amplitudes, human perception, and global and local recovery.
    \bulletitem{Compression:} We examine global recovery and code images recovered under different levels of compression. 
    % We also examine the impact of compression used by popular social media and video sharing sites.
    \bulletitem{Temporal Analysis Window Size:} We examine the effect of different window sizes for analysis.
    \bulletitem{Motion:} This includes analysis and ablations for our filtering strategies applied to motion of different speeds and using analysis windows of different sizes.
    \bulletitem{Photon Shot Noise:} We examine the impact of uncoded light on SNR, and the relationship between noise and quantization.
    \bulletitem{Quantization:} We examine the impacts of quantization and gamma encoding on \nci{}.
\end{itemize}
These experiments provide a useful reference for predicting how reliable \nci{} will be under different conditions. We predict the SNR of \nci{} parameterized by code signal strength, amount of uncoded light, temporal analysis window size, and spatial downsample factor assuming photon and read noise sources in Section \ref{sec:snr} of the appendix.
% In practice, we found that \nci{} is robust to a wide range of realistic conditions, but there are limitations. We discuss these more in Section \ref{sec:limit} and refer to our supplemental for full details.

% These experiments are especially informative regarding the limits of \nci{}, and we  them to better understand

\subsection{Real Experiments}
\label{sec:realexperiments}
\aadd{Table ~\ref{tab:setup} in the appendix shows the layout for several of our captured scenes. We demonstrate \nci{} in \nscenes{} different real-world scenes covering a wide range of possible scenarios, including a speech setting in a lecture hall (\FigRef{fig:teaser}), a conference room presentation (\FigRef{fig:lightingdecomposition}), a television interview setting (\FigRef{fig:interview}), an outdoor scene dominated by sunlight (\FigRef{fig:outdoor_setup}), and a public event (with permission of the recorded speakers) where a real audience of over 100 people were present who did not notice the coded illumination (\FigRef{fig:motionweight})}. Unless stated otherwise, videos were recorded using the default compression settings of whatever device was used for capture. Figure \ref{fig:relit} shows an example of using \nci{} for scene relighting. 
\revcomment{Substantial shortening}
\padd{Figure \ref{fig:dtw} contains examples of analyzing different kinds of temporal tampering with video compressed by popular video hosting websites.
We encourage readers to explore our supplemental material and web gallery, which include:
\begin{itemize}
    \item Additional details for each experiment
    \item Additional results, including a dark-skinned human subject, a deepfake example, and different compression levels
    % \item Example using our method for scene relighting
    \item Analysis of predicted flicker sensitivity likelihood from using coded illumination
    \item Ablation of our patch-selection algorithm on the outdoor scene
    \item Ablation of bilateral filtering for removing temporal transients from motion and camera flash
    % \item Captured and manipulated videos for reference
\end{itemize}}

% /////////////////////////////////////////////////% /////////////////////////////////////////////////

\subsection{Informed Adversarial Attacks}
\label{sec:adversarial}
\aadd{Our supplemental material also examines various strategies that an informed adversary--meaning one that understands \nci{} and knows illumination has been coded--might use to hide manipulation in a video. While it is not possible to anticipate all potential strategies, we show that, at least for several informed attack strategies, attempts to estimate and fake some aspect of coded illumination tend to leave detectable artifacts. Most of these artifacts happen because the code signal acts as a carrier for information from the original video, so even small errors in a code estimate tend to either leave traces of the original video or make fake content inconsistent with the rest of the scene. 
\revcomment{\textcolor{black}{Moved to supplemental, replaced with summary}}
However, as these artifacts can be different for different attacks, and adversaries may use new strategies with different tradeoffs over time, we stress the importance of considering evidence in-context as part of a broad forensic analysis, ideally involving a human analyst, in practice. We also examine one very limited type of manipulation (reflectance-only edits) that is especially difficult to detect with \nci{}, which we also discuss in the limitations section of this paper.}

\section{Discussion}
\label{sec:discussion}

\subsection{Detecting Temporal Manipulation}
\label{sec:discussiontemporal}
% The most robust use case for \nci{} seems to be the detection of temporal manipulation. This is because a
Aggregating over high-SNR regions in a video tends to produce a very reliable global alignment with code signals, which makes detecting temporal manipulation of video very robust. Our supplemental experiments confirm that this robustness extends to very high levels of compression, and, when our bilateral filtering strategy is used, large amounts of scene motion. We attribute this robustness to the fact that our code is spread across many temporal frequencies and pixels.

% extreme levels  bilateral filtering extends this robustness to scenes with significant motion, and that alignment needed to detect temporal manipulation is fairly robust to scene motion when we use our temporal bilateral filtering, and that reliable temporal alignment is possible even at fairly extreme levels of compression, even when such compression renders recovered code images difficult to interpret. 

% Our compression experiments reinforce this point, as temporal manipulation can still be detected even when compression renders code images nearly uninterpretable. 

Our interview scene shows a particularly compelling use case for \nci{}, where temporal analysis is able to reveal a detailed description of edits made even without access to a reference recording of the scene. In this sense, we can think of \nci{} as offering some of the same benefits as a reference recording of the scene, but with a few advantages:
\begin{itemize}
\item \nci{} does not require any noticeable change or addition to the scene (e.g., the placement of an additional camera).
\item \nci{} can cover large spaces for extended time periods without the corresponding cost of cameras and storage. 
\item Registering video in time with a code used in \nci{} is much simpler than searching through a reference recording, as it does not require finding correspondences between different viewpoints of a scene.
\end{itemize}
We can think of \nci{} as a way to subtly timestamp any video captured in a particular environment. The cost of adding \nci{} to a scene is almost negligible, making it an especially appealing way to add long-term protection in environments that frequently host likely targets of disinformation.   

% especially compared to the cost of adding CCTV cameras   it a practical protection to add to large spaces with likely targets and uncontrolled cameras (e.g., diplomatic gatherings).

% Our outdoor scene demonstrates an interesting case where coded fluctuations are largely uncorrelated with average are relatively small compared to variations due to shot noise. In this case, a simple aglobal average of each frame will fail to align with the illumination code due to high noise levels in pixels with high amounts of uncoded light from the Sun. In this case, knowing the code makes it relatively simple to find regions of the video with higher SNR that can be used for alignment, but without knowing the code, these regions are hard to find.
% this would be
%Notably, the ability to detect temporal manipulation makes deep face pupetting particularly difficult.

\subsection{Detecting Spatial Manipulation}
\nci{} is effective at detecting spatial manipulation in many scenarios, particularly when most of the illumination in a scene is coded.
For edits made by an uninformed adversary (meaning one that manipulates a video without knowing that \nci{} was used in the original scene), spatial manipulation is straightforward to detect in regions of the scene that are exposed to coded illumination. However, as code images rely on spatially localized information, we find that the reliability of spatial analysis degrades much faster than temporal analysis with extreme compression, or high levels of uncoded light or scene motion. We discuss these limits more in \Sectn{\ref{sec:limit}}.

\subsection{Limitations}
\label{sec:limit}
% \nci{} offers many advantages, but it is not a complete solution to the forensics problem. 
\nci{} has many advantages, but it is not a perfect solution to the forensics problem. 
It requires a physical intervention in the scene being protected, and while it creates an advantage for forensic analysis, it cannot guarantee with complete certainty that manipulation will be detected. With this in mind, analysts should always consider factors that can make analysis less conclusive.

\subsubsection{Motion}
Our code analysis relies on observations over a time window, which can create problems with moving objects. Bilateral filtering and \transientfiltered{} code images help limit the impact of motion, but scenes with significant amounts of constant motion that cannot easily be tracked or stabilized still pose a limitation for \nci{}. 

% and provides a substantial improvement on the former but its limited degrees of control make it tough to remove all transients without degrading the code signal. We believe that more advanced filtering techniques can solve this problem, such as ones guided by the ground truth code signal.

% Our supplemental material includes a detailed look at the impact of different frequencies and amplitudes of motion. We saw in our experimental setups that code images degrade with motion. Recalling Section \ref{sec:motion}, motion causes two key problems: the addition of temporal transients and motion blur in code images once these transients are removed. The bilateral filter provides a substantial improvement on the former but its limited degrees of control make it tough to remove all transients without degrading the code signal. We believe that more advanced filtering techniques can solve this problem, such as ones guided by the ground truth code signal.

% However, removing motion blur in code images is fundamentally limited by motion due to the temporal nature of the coding used by \nci{}.

\subsubsection{Regions with Low Coded Variation}
\nci{} is only effective when coded variation is actually present in a video. This means that for parts of a scene that do not reflect coded light (e.g., low reflectance regions or those in the shadow of all coded sources), or for over-exposed regions with saturated pixels, \nci{} analysis may be inconclusive.

\subsubsection{Regions with High Uncoded Light}
High amounts of uncoded light result in high levels of shot noise, which reduces the SNR of observed coded variations. In our outdoor experiment, shot noise from sunlight was strong enough to cause global alignment with our code signal to fail. However, knowledge of the illumination code let us identify local regions with higher SNR, which we used to align the video and recover a code image. Here, the low SNR of pixels with high shot noise can limit our analysis in parts of the scene. However, this may also offer some added protection against informed adversaries, as it makes coded variation less distinguishable from spurious correlation of noise. 

\subsubsection{Compression}
Our supplemental material includes experiments that examine a wide range of video compression. We verified that our analysis is largely robust to compression from various popular video hosting websites (see \Fig{} \ref{fig:dtw} and our supplemental material). Our supplemental material also includes an analysis of different compression ratios on simulated video with carefully controlled characteristics. We found that spatial and temporal analysis are both fairly robust to common compression levels, but spatial analysis (code images) degrades faster than temporal analysis under extreme levels of compression. We also analyze the effects of gamma encoding and 8-bit quantization in our supplemental material, but find that in practice, these effects are minimized by the presence of photon shot noise.

\subsubsection{Future Informed Attacks \& Reflectance-Only Manipulation}
\label{sec:limitinformed}
% Our experiments and analysis suggest that this is at least difficult to do in general, but the risk of such attacks should be tracked and reassessed over time. 
% In \Sectn{\ref{sec:reflectanceonlyattack}}, we identified a strategy for manipulating the reflectance of surfaces that is difficult to detect with \nci{}. 
While \nci{} appears to be robust to most basic informed attacks, we identified one type of manipulation that is especially difficult to detect with \nci{}; that is, manipulation that changes the reflectance in a scene without changing its geometry or remapping time. Such changes can be done with a per-pixel, per-channel gain, which results in a video that is otherwise indistinguishable from that of a scene with different values of $\reflectance{}$. While such changes are limited, they are important to consider.  

While our experiments and analysis in the supplemental suggest that hiding more general types of manipulation is at least very difficult, there is always a possibility that informed adversaries will find new ways to circumvent \nci{}. This risk should be tracked and reassessed over time. Additional research on \nci{} analysis can also help alleviate this risk, as we describe in \Sectn{\ref{sec:futurework}}.

\subsection{Future Work}
\label{sec:futurework}
Our work is a promising first step in exploring the use of coded illumination in forensics, but there is much room for future work.

\subsubsection{Data-Driven \nci{} Analysis}
Our work in this paper focused on understanding \nci{} as an imaging modality and understanding how the manifold reduction and information asymmetry that it creates can benefit forensic analysis. In the longer term, we believe that having a human participate in forensic analysis is important for incorporating context, but there could be many ways to supplement \nci{} with data-driven priors. The training and use of these priors is an exciting direction for future work.

\subsubsection{Coded Source Distribution}
Our experiments only use a small number of illumination codes, and we do not explore the arrangement of coded sources in great detail. There may be strategic ways to distribute coded sources in a scene for increased robustness.

% \subsubsection{Structured Light Sources}
% Using a projector as a coded source allows us to make our code vary spatially in potentially complex ways. In the extreme case, each projected pixel can have a distinct code, which would make estimating them significantly harder.

\subsubsection{Adversarial Attacks}
As with other research topics related to forensics and security, there is always a threat of new and more advanced adversarial attacks. Understanding these attacks is the first step to countering them.
% Our work introduces \nci{} as a promising strategy for protecting the contents of a particular scene from use in manipulated video. As a new imag 
% The forensics problem is one that can never be solved completely.

\subsubsection{Deployment}
Our paper demonstrates the potential value of \nci{}, but many questions remain about how best to deploy it. In particular, how to manage access to illumination codes and how to securely assess and present evidence of manipulation are both important practical questions that affect the potential long-term impact of \nci{}.   

\section{Conclusion}
\label{sec:conclusion}
Our work introduces noise-coded illumination (\nci{}) as a novel forensic strategy that helps protect content in a particular physical space. 
% a particular environment from being used in specific environments from  from way to help protect content in a physical space from   
% analysis of video captured in a specific environment.
% help forensic analysts identify manipulations of video captured in a protected environment. 
Our approach creates an information asymmetry by using randomized illumination codes that resemble noise. It also makes manipulations easier to detect by reducing the manifold of plausible videos. Our approach is inexpensive, simple to implement, and unnoticeable to most observers. We validate our approach on a variety of realistic scenes and subjects, and analyze the effects of factors including compression, motion, photon noise, and window size. Our results suggest that \nci{} is a promising new approach to video forensics with significant potential for future work.

% \subsubsection{Adversarial Attacks}
% As with other research topics related to forensics and security, there is always a threat of new and more advanced adversarial attacks. Understanding these attacks is the first step to countering them.
% % Our work introduces \nci{} as a promising strategy for protecting the contents of a particular scene from use in manipulated video. As a new imag 
% % The forensics problem is one that can never be solved completely.

% Our work does not solve the forensics problem, but it offers a new and powerful approach that shows significant promise for protecting sensitive scenes. We believe this work opens an exciting new research direction with potential for significant impact on video forensics.

% more diagnostic result for detecting temporal tamper (was ACC this much etc)
% auto detect spatial tamper (lighting inconsistency)
% noise analysis for detecting tampers, using original lighting conditions
% better robustness to motion,compression,outdoors
% add noise to lights
 % infrared lights

\begin{acks}
We thank all those who volunteered to be subjects in our test scenes. This work was supported in part by an NDSEG fellowship to P.M., and the Pioneer Centre for AI, DNRF grant number P1.
\end{acks}

\bibliographystyle{ACM-Reference-Format}
\bibliography{main}


\begin{thebibliography}{62}

%%% the Web address.

\ifx \showCODEN    \undefined \def \showCODEN     #1{\unskip}     \fi
\ifx \showISBNx    \undefined \def \showISBNx     #1{\unskip}     \fi
\ifx \showISBNxiii \undefined \def \showISBNxiii  #1{\unskip}     \fi
\ifx \showISSN     \undefined \def \showISSN      #1{\unskip}     \fi
\ifx \showLCCN     \undefined \def \showLCCN      #1{\unskip}     \fi
\ifx \shownote     \undefined \def \shownote      #1{#1}          \fi
\ifx \showarticletitle \undefined \def \showarticletitle #1{#1}   \fi
\ifx \showURL      \undefined \def \showURL       {\relax}        \fi
% invisible to TeX
\providecommand\bibfield[2]{#2}
\providecommand\bibinfo[2]{#2}
\providecommand\natexlab[1]{#1}
\providecommand\showeprint[2][]{arXiv:#2}

\bibitem[Agarwal et~al\mbox{.}(2020)]%
        {agarwal2020detecting}
\bibfield{author}{\bibinfo{person}{Shruti Agarwal}, \bibinfo{person}{Hany
  Farid}, \bibinfo{person}{Tarek El-Gaaly}, {and} \bibinfo{person}{Ser-Nam
  Lim}.} \bibinfo{year}{2020}\natexlab{}.
\newblock \showarticletitle{Detecting Deep-Fake Videos from Appearance and
  Behavior}. In \bibinfo{booktitle}{\emph{2020 IEEE International Workshop on
  Information Forensics and Security (WIFS)}}. \bibinfo{pages}{1--6}.
\newblock
\href{https://doi.org/10.1109/WIFS49906.2020.9360904}{doi:\nolinkurl{10.1109/WIFS49906.2020.9360904}}


\bibitem[Ahn et~al\mbox{.}(2021)]%
        {kalstructuredlight}
\bibfield{author}{\bibinfo{person}{Byeongjoo Ahn}, \bibinfo{person}{Ioannis
  Gkioulekas}, {and} \bibinfo{person}{Aswin~C. Sankaranarayanan}.}
  \bibinfo{year}{2021}\natexlab{}.
\newblock \showarticletitle{Kaleidoscopic structured light}.
\newblock \bibinfo{journal}{\emph{ACM Trans. Graph.}} \bibinfo{volume}{40},
  \bibinfo{number}{6}, Article \bibinfo{articleno}{214} (\bibinfo{date}{dec}
  \bibinfo{year}{2021}), \bibinfo{numpages}{15}~pages.
\newblock
\showISSN{0730-0301}
\href{https://doi.org/10.1145/3478513.3480524}{doi:\nolinkurl{10.1145/3478513.3480524}}


\bibitem[Asikuzzaman and Pickering(2018)]%
        {videowatermark}
\bibfield{author}{\bibinfo{person}{Md. Asikuzzaman} {and}
  \bibinfo{person}{Mark~R. Pickering}.} \bibinfo{year}{2018}\natexlab{}.
\newblock \showarticletitle{An Overview of Digital Video Watermarking}.
\newblock \bibinfo{journal}{\emph{IEEE Transactions on Circuits and Systems for
  Video Technology}} \bibinfo{volume}{28}, \bibinfo{number}{9}
  (\bibinfo{year}{2018}), \bibinfo{pages}{2131--2153}.
\newblock
\href{https://doi.org/10.1109/TCSVT.2017.2712162}{doi:\nolinkurl{10.1109/TCSVT.2017.2712162}}


\bibitem[Bauder and Woodward(2018)]%
        {AcostaFake}
\bibfield{author}{\bibinfo{person}{David Bauder} {and} \bibinfo{person}{Calvin
  Woodward}.} \bibinfo{year}{2018}\natexlab{}.
\newblock \showarticletitle{Expert: Acosta video distributed by White House was
  doctored}.
\newblock \bibinfo{journal}{\emph{AP News}} (\bibinfo{date}{8 November}
  \bibinfo{year}{2018}).
\newblock
\urldef\tempurl%
\url{https://apnews.com/article/c575bd1cc3b1456cb3057ef670c7fe2a}
\showURL{%
\tempurl}


\bibitem[Bobby~Allyn(2022)]%
        {zelensky}
\bibfield{author}{\bibinfo{person}{National Public~Radio Bobby~Allyn}.}
  \bibinfo{year}{2022}\natexlab{}.
\newblock \bibinfo{booktitle}{\emph{Deepfake video of Zelenskyy could be 'tip
  of the iceberg' in info war, experts warn}}.
\newblock
\urldef\tempurl%
\url{https://www.npr.org/2022/03/16/1087062648/deepfake-video-zelenskyy-experts-war-manipulation-ukraine-russia}
\showURL{%
\tempurl}


\bibitem[Bodington et~al\mbox{.}(2016)]%
        {flicker}
\bibfield{author}{\bibinfo{person}{D Bodington}, \bibinfo{person}{A Bierman},
  {and} \bibinfo{person}{N Narendran}.} \bibinfo{year}{2016}\natexlab{}.
\newblock \showarticletitle{A flicker perception metric}.
\newblock \bibinfo{journal}{\emph{Lighting Research \& Technology}}
  \bibinfo{volume}{48}, \bibinfo{number}{5} (\bibinfo{year}{2016}),
  \bibinfo{pages}{624--641}.
\newblock
\href{https://doi.org/10.1177/1477153515581006}{doi:\nolinkurl{10.1177/1477153515581006}}
\showeprint{https://doi.org/10.1177/1477153515581006}


\bibitem[Bui et~al\mbox{.}(2023)]%
        {rosteal}
\bibfield{author}{\bibinfo{person}{Tu Bui}, \bibinfo{person}{Shruti Agarwal},
  \bibinfo{person}{Ning Yu}, {and} \bibinfo{person}{John Collomosse}.}
  \bibinfo{year}{2023}\natexlab{}.
\newblock \showarticletitle{RoSteALS: Robust Steganography Using Autoencoder
  Latent Space}. In \bibinfo{booktitle}{\emph{Proceedings of the IEEE/CVF
  Conference on Computer Vision and Pattern Recognition (CVPR) Workshops}}.
  \bibinfo{pages}{933--942}.
\newblock


\bibitem[Cozzolino et~al\mbox{.}(2023)]%
        {avpoi}
\bibfield{author}{\bibinfo{person}{Davide Cozzolino},
  \bibinfo{person}{Alessandro Pianese}, \bibinfo{person}{Matthias Nie{\ss}ner},
  {and} \bibinfo{person}{Luisa Verdoliva}.} \bibinfo{year}{2023}\natexlab{}.
\newblock \showarticletitle{Audio-Visual Person-of-Interest DeepFake
  Detection}. In \bibinfo{booktitle}{\emph{Proceedings of the IEEE/CVF
  Conference on Computer Vision and Pattern Recognition (CVPR) Workshops}}.
  \bibinfo{pages}{943--952}.
\newblock


\bibitem[Cozzolino et~al\mbox{.}(2019)]%
        {Verdoliva_2019_CVPR_Workshops}
\bibfield{author}{\bibinfo{person}{Davide Cozzolino}, \bibinfo{person}{Giovanni
  Poggi}, {and} \bibinfo{person}{Luisa Verdoliva}.}
  \bibinfo{year}{2019}\natexlab{}.
\newblock \showarticletitle{Extracting camera-based fingerprints for video
  forensics}. In \bibinfo{booktitle}{\emph{Proceedings of the IEEE/CVF
  Conference on Computer Vision and Pattern Recognition (CVPR) Workshops}}.
\newblock


\bibitem[Davis et~al\mbox{.}(2017)]%
        {VisVibPAMI}
\bibfield{author}{\bibinfo{person}{Abe Davis}, \bibinfo{person}{Katherine~L.
  Bouman}, \bibinfo{person}{Justin~G. Chen}, \bibinfo{person}{Michael
  Rubinstein}, \bibinfo{person}{Fredo Durand}, {and}
  \bibinfo{person}{William~T. Freeman}.} \bibinfo{year}{2017}\natexlab{}.
\newblock \showarticletitle{Visual Vibrometry: Estimating Material Properties
  from Small Motions in Video}.
\newblock \bibinfo{journal}{\emph{IEEE Transactions on Pattern Analysis and
  Machine Intelligence}} \bibinfo{volume}{39}, \bibinfo{number}{4}
  (\bibinfo{date}{April} \bibinfo{year}{2017}), \bibinfo{pages}{732--745}.
\newblock
\showISSN{0162-8828}
\href{https://doi.org/10.1109/TPAMI.2016.2622271}{doi:\nolinkurl{10.1109/TPAMI.2016.2622271}}


\bibitem[Davis et~al\mbox{.}(2015)]%
        {DavisISMB}
\bibfield{author}{\bibinfo{person}{Abe Davis}, \bibinfo{person}{Justin~G.
  Chen}, {and} \bibinfo{person}{Fr{\'e}do Durand}.}
  \bibinfo{year}{2015}\natexlab{}.
\newblock \showarticletitle{Image-space Modal Bases for Plausible Manipulation
  of Objects in Video}.
\newblock \bibinfo{journal}{\emph{ACM Trans. Graph.}} \bibinfo{volume}{34},
  \bibinfo{number}{6}, Article \bibinfo{articleno}{239} (\bibinfo{date}{Oct.}
  \bibinfo{year}{2015}), \bibinfo{numpages}{7}~pages.
\newblock
\showISSN{0730-0301}
\href{https://doi.org/10.1145/2816795.2818095}{doi:\nolinkurl{10.1145/2816795.2818095}}


\bibitem[Davis et~al\mbox{.}(2014)]%
        {VisMic}
\bibfield{author}{\bibinfo{person}{Abe Davis}, \bibinfo{person}{Michael
  Rubinstein}, \bibinfo{person}{Neal Wadhwa}, \bibinfo{person}{Gautham~J.
  Mysore}, \bibinfo{person}{Fr{\'e}do Durand}, {and}
  \bibinfo{person}{William~T. Freeman}.} \bibinfo{year}{2014}\natexlab{}.
\newblock \showarticletitle{The Visual Microphone: Passive Recovery of Sound
  from Video}.
\newblock \bibinfo{journal}{\emph{ACM Trans. Graph.}} \bibinfo{volume}{33},
  \bibinfo{number}{4}, Article \bibinfo{articleno}{79} (\bibinfo{date}{July}
  \bibinfo{year}{2014}), \bibinfo{numpages}{10}~pages.
\newblock
\showISSN{0730-0301}
\href{https://doi.org/10.1145/2601097.2601119}{doi:\nolinkurl{10.1145/2601097.2601119}}


\bibitem[de~Lange~Dzn(1961)]%
        {de1961eye}
\bibfield{author}{\bibinfo{person}{H de Lange~Dzn}.}
  \bibinfo{year}{1961}\natexlab{}.
\newblock \showarticletitle{Eye’s response at flicker fusion to square-wave
  modulation of a test field surrounded by a large steady field of equal mean
  luminance}.
\newblock \bibinfo{journal}{\emph{JOSA}} \bibinfo{volume}{51},
  \bibinfo{number}{4} (\bibinfo{year}{1961}), \bibinfo{pages}{415--421}.
\newblock


\bibitem[Eisemann and Durand(2004)]%
        {fnf}
\bibfield{author}{\bibinfo{person}{Elmar Eisemann} {and}
  \bibinfo{person}{Fr{\'e}do Durand}.} \bibinfo{year}{2004}\natexlab{}.
\newblock \showarticletitle{Flash photography enhancement via intrinsic
  relighting}.
\newblock \bibinfo{journal}{\emph{ACM transactions on graphics (TOG)}}
  \bibinfo{volume}{23}, \bibinfo{number}{3} (\bibinfo{year}{2004}),
  \bibinfo{pages}{673--678}.
\newblock


\bibitem[{Espressif}(2016)]%
        {esp32}
\bibfield{author}{\bibinfo{person}{{Espressif}}.}
  \bibinfo{year}{2016}\natexlab{}.
\newblock \bibinfo{title}{{ESP32: Wi-Fi \& Bluetooth Microcontroller}}.
\newblock
  \bibinfo{howpublished}{\url{https://www.espressif.com/en/products/socs/esp32}}.
\newblock


\bibitem[Feng et~al\mbox{.}(2023)]%
        {audvid}
\bibfield{author}{\bibinfo{person}{Chao Feng}, \bibinfo{person}{Ziyang Chen},
  {and} \bibinfo{person}{Andrew Owens}.} \bibinfo{year}{2023}\natexlab{}.
\newblock \showarticletitle{Self-Supervised Video Forensics by Audio-Visual
  Anomaly Detection}. In \bibinfo{booktitle}{\emph{Proceedings of the IEEE/CVF
  Conference on Computer Vision and Pattern Recognition (CVPR)}}.
  \bibinfo{pages}{10491--10503}.
\newblock


\bibitem[Freedman et~al\mbox{.}(2014)]%
        {SRATOF}
\bibfield{author}{\bibinfo{person}{Daniel Freedman}, \bibinfo{person}{Yoni
  Smolin}, \bibinfo{person}{Eyal Krupka}, \bibinfo{person}{Ido Leichter}, {and}
  \bibinfo{person}{Mirko Schmidt}.} \bibinfo{year}{2014}\natexlab{}.
\newblock \showarticletitle{SRA: Fast Removal of General Multipath for ToF
  Sensors}. In \bibinfo{booktitle}{\emph{Computer Vision -- ECCV 2014}},
  \bibfield{editor}{\bibinfo{person}{David Fleet}, \bibinfo{person}{Tomas
  Pajdla}, \bibinfo{person}{Bernt Schiele}, {and} \bibinfo{person}{Tinne
  Tuytelaars}} (Eds.). \bibinfo{publisher}{Springer International Publishing},
  \bibinfo{address}{Cham}, \bibinfo{pages}{234--249}.
\newblock
\showISBNx{978-3-319-10590-1}


\bibitem[Gerstner and Farid(2022)]%
        {farid}
\bibfield{author}{\bibinfo{person}{Candice~R Gerstner} {and}
  \bibinfo{person}{Hany Farid}.} \bibinfo{year}{2022}\natexlab{}.
\newblock \showarticletitle{Detecting real-time deep-fake videos using active
  illumination}. In \bibinfo{booktitle}{\emph{Proceedings of the IEEE/CVF
  Conference on Computer Vision and Pattern Recognition}}.
  \bibinfo{pages}{53--60}.
\newblock


\bibitem[Hannah~Denham(2020)]%
        {PelosiSlur}
\bibfield{author}{\bibinfo{person}{The Washington~Post Hannah~Denham}.}
  \bibinfo{year}{2020}\natexlab{}.
\newblock \bibinfo{booktitle}{\emph{Another fake video of Pelosi goes viral on
  Facebook}}.
\newblock
\urldef\tempurl%
\url{https://www.washingtonpost.com/technology/2020/08/03/nancy-pelosi-fake-video-facebook/}
\showURL{%
\tempurl}


\bibitem[Hartung and Girod(1998)]%
        {watermarknoise}
\bibfield{author}{\bibinfo{person}{Frank Hartung} {and} \bibinfo{person}{Bernd
  Girod}.} \bibinfo{year}{1998}\natexlab{}.
\newblock \showarticletitle{Watermarking of uncompressed and compressed video}.
\newblock \bibinfo{journal}{\emph{Signal Processing}} \bibinfo{volume}{66},
  \bibinfo{number}{3} (\bibinfo{year}{1998}), \bibinfo{pages}{283--301}.
\newblock
\showISSN{0165-1684}
\href{https://doi.org/10.1016/S0165-1684(98)00011-5}{doi:\nolinkurl{10.1016/S0165-1684(98)00011-5}}


\bibitem[Heide et~al\mbox{.}(2013)]%
        {LightInFlightFelix}
\bibfield{author}{\bibinfo{person}{Felix Heide}, \bibinfo{person}{Matthias~B.
  Hullin}, \bibinfo{person}{James Gregson}, {and} \bibinfo{person}{Wolfgang
  Heidrich}.} \bibinfo{year}{2013}\natexlab{}.
\newblock \showarticletitle{Low-Budget Transient Imaging Using Photonic Mixer
  Devices}.
\newblock \bibinfo{journal}{\emph{ACM Trans. Graph.}} \bibinfo{volume}{32},
  \bibinfo{number}{4}, Article \bibinfo{articleno}{45} (\bibinfo{date}{July}
  \bibinfo{year}{2013}), \bibinfo{numpages}{10}~pages.
\newblock
\showISSN{0730-0301}
\href{https://doi.org/10.1145/2461912.2461945}{doi:\nolinkurl{10.1145/2461912.2461945}}


\bibitem[Heide et~al\mbox{.}(2019)]%
        {NLOSWOccluders}
\bibfield{author}{\bibinfo{person}{Felix Heide}, \bibinfo{person}{Matthew
  O’Toole}, \bibinfo{person}{Kai Zang}, \bibinfo{person}{David~B. Lindell},
  \bibinfo{person}{Steven Diamond}, {and} \bibinfo{person}{Gordon Wetzstein}.}
  \bibinfo{year}{2019}\natexlab{}.
\newblock \showarticletitle{Non-line-of-sight Imaging with Partial Occluders
  and Surface Normals}.
\newblock \bibinfo{journal}{\emph{ACM Trans. Graph.}} \bibinfo{volume}{38},
  \bibinfo{number}{3}, Article \bibinfo{articleno}{22} (\bibinfo{date}{May}
  \bibinfo{year}{2019}), \bibinfo{numpages}{10}~pages.
\newblock
\showISSN{0730-0301}
\href{https://doi.org/10.1145/3269977}{doi:\nolinkurl{10.1145/3269977}}


\bibitem[Huang et~al\mbox{.}(2022)]%
        {cmua}
\bibfield{author}{\bibinfo{person}{Hao Huang}, \bibinfo{person}{Yongtao Wang},
  \bibinfo{person}{Zhaoyu Chen}, \bibinfo{person}{Yuze Zhang},
  \bibinfo{person}{Yuheng Li}, \bibinfo{person}{Zhi Tang}, \bibinfo{person}{Wei
  Chu}, \bibinfo{person}{Jingdong Chen}, \bibinfo{person}{Weisi Lin}, {and}
  \bibinfo{person}{Kai-Kuang Ma}.} \bibinfo{year}{2022}\natexlab{}.
\newblock \showarticletitle{Cmua-watermark: A cross-model universal adversarial
  watermark for combating deepfakes}. In \bibinfo{booktitle}{\emph{Proceedings
  of the AAAI Conference on Artificial Intelligence}},
  Vol.~\bibinfo{volume}{36}. \bibinfo{pages}{989--997}.
\newblock


\bibitem[Huh et~al\mbox{.}(2018)]%
        {huh2018fighting}
\bibfield{author}{\bibinfo{person}{Minyoung Huh}, \bibinfo{person}{Andrew Liu},
  \bibinfo{person}{Andrew Owens}, {and} \bibinfo{person}{Alexei~A. Efros}.}
  \bibinfo{year}{2018}\natexlab{}.
\newblock \showarticletitle{Fighting Fake News: Image Splice Detection via
  Learned Self-Consistency}. In \bibinfo{booktitle}{\emph{Computer Vision –
  ECCV 2018: 15th European Conference, Munich, Germany, September 8-14, 2018,
  Proceedings, Part XI}} (Munich, Germany).
  \bibinfo{publisher}{Springer-Verlag}, \bibinfo{address}{Berlin, Heidelberg},
  \bibinfo{pages}{106–124}.
\newblock
\showISBNx{978-3-030-01251-9}
\href{https://doi.org/10.1007/978-3-030-01252-6_7}{doi:\nolinkurl{10.1007/978-3-030-01252-6_7}}


\bibitem[{IEEE}(2016)]%
        {DSSS_IEEE802p11}
\bibfield{author}{\bibinfo{person}{{IEEE}}.} \bibinfo{year}{2016}\natexlab{}.
\newblock \showarticletitle{IEEE Standard for Information
  technology—Telecommunications and information exchange between systems
  Local and metropolitan area networks—Specific requirements - Part 11:
  Wireless LAN Medium Access Control (MAC) and Physical Layer (PHY)
  Specifications}.
\newblock \bibinfo{journal}{\emph{IEEE Std 802.11-2016 (Revision of IEEE Std
  802.11-2012)}} (\bibinfo{year}{2016}), \bibinfo{pages}{1--3534}.
\newblock
\href{https://doi.org/10.1109/IEEESTD.2016.7786995}{doi:\nolinkurl{10.1109/IEEESTD.2016.7786995}}


\bibitem[Kee et~al\mbox{.}(2013)]%
        {faridshadows}
\bibfield{author}{\bibinfo{person}{Eric Kee}, \bibinfo{person}{James~F.
  O'Brien}, {and} \bibinfo{person}{Hany Farid}.}
  \bibinfo{year}{2013}\natexlab{}.
\newblock \showarticletitle{Exposing photo manipulation with inconsistent
  shadows}.
\newblock \bibinfo{journal}{\emph{ACM Trans. Graph.}} \bibinfo{volume}{32},
  \bibinfo{number}{3}, Article \bibinfo{articleno}{28} (\bibinfo{date}{jul}
  \bibinfo{year}{2013}), \bibinfo{numpages}{12}~pages.
\newblock
\showISSN{0730-0301}
\href{https://doi.org/10.1145/2487228.2487236}{doi:\nolinkurl{10.1145/2487228.2487236}}


\bibitem[Kee et~al\mbox{.}(2014)]%
        {faridshading}
\bibfield{author}{\bibinfo{person}{Eric Kee}, \bibinfo{person}{James~F.
  O'brien}, {and} \bibinfo{person}{Hany Farid}.}
  \bibinfo{year}{2014}\natexlab{}.
\newblock \showarticletitle{Exposing Photo Manipulation from Shading and
  Shadows}.
\newblock \bibinfo{journal}{\emph{ACM Trans. Graph.}} \bibinfo{volume}{33},
  \bibinfo{number}{5}, Article \bibinfo{articleno}{165} (\bibinfo{date}{sep}
  \bibinfo{year}{2014}), \bibinfo{numpages}{21}~pages.
\newblock
\showISSN{0730-0301}
\href{https://doi.org/10.1145/2629646}{doi:\nolinkurl{10.1145/2629646}}


\bibitem[Kelly(1961)]%
        {kelly1961visual}
\bibfield{author}{\bibinfo{person}{DH Kelly}.} \bibinfo{year}{1961}\natexlab{}.
\newblock \showarticletitle{Visual responses to time-dependent stimuli.* i.
  amplitude sensitivity measurements}.
\newblock \bibinfo{journal}{\emph{JOSA}} \bibinfo{volume}{51},
  \bibinfo{number}{4} (\bibinfo{year}{1961}), \bibinfo{pages}{422--429}.
\newblock


\bibitem[{Korus} and {Memon}(2019)]%
        {neuralimagingpipeline}
\bibfield{author}{\bibinfo{person}{P. {Korus}} {and} \bibinfo{person}{N.
  {Memon}}.} \bibinfo{year}{2019}\natexlab{}.
\newblock \showarticletitle{Content Authentication for Neural Imaging
  Pipelines: End-To-End Optimization of Photo Provenance in Complex
  Distribution Channels}. In \bibinfo{booktitle}{\emph{2019 IEEE/CVF Conference
  on Computer Vision and Pattern Recognition (CVPR)}}.
  \bibinfo{pages}{8613--8621}.
\newblock
\href{https://doi.org/10.1109/CVPR.2019.00882}{doi:\nolinkurl{10.1109/CVPR.2019.00882}}


\bibitem[{Lei Xiao} et~al\mbox{.}(2015)]%
        {7298851}
\bibfield{author}{\bibinfo{person}{{Lei Xiao}}, \bibinfo{person}{F. {Heide}},
  \bibinfo{person}{M. {O'Toole}}, \bibinfo{person}{A. {Kolb}},
  \bibinfo{person}{M.~B. {Hullin}}, \bibinfo{person}{K. {Kutulakos}}, {and}
  \bibinfo{person}{W. {Heidrich}}.} \bibinfo{year}{2015}\natexlab{}.
\newblock \showarticletitle{Defocus deblurring and superresolution for
  time-of-flight depth cameras}. In \bibinfo{booktitle}{\emph{2015 IEEE
  Conference on Computer Vision and Pattern Recognition (CVPR)}}.
  \bibinfo{pages}{2376--2384}.
\newblock
\href{https://doi.org/10.1109/CVPR.2015.7298851}{doi:\nolinkurl{10.1109/CVPR.2015.7298851}}


\bibitem[Li and Stuber(2006)]%
        {freqmultiplex}
\bibfield{author}{\bibinfo{person}{Ye~Geoffrey Li} {and}
  \bibinfo{person}{Gordon~L Stuber}.} \bibinfo{year}{2006}\natexlab{}.
\newblock \bibinfo{booktitle}{\emph{Orthogonal frequency division multiplexing
  for wireless communications}}.
\newblock \bibinfo{publisher}{Springer Science \& Business Media}.
\newblock


\bibitem[{Lin} et~al\mbox{.}(2017)]%
        {FDomainTransient}
\bibfield{author}{\bibinfo{person}{J. {Lin}}, \bibinfo{person}{Y. {Liu}},
  \bibinfo{person}{J. {Suo}}, {and} \bibinfo{person}{Q. {Dai}}.}
  \bibinfo{year}{2017}\natexlab{}.
\newblock \showarticletitle{Frequency-Domain Transient Imaging}.
\newblock \bibinfo{journal}{\emph{IEEE Transactions on Pattern Analysis and
  Machine Intelligence}} \bibinfo{volume}{39}, \bibinfo{number}{5}
  (\bibinfo{year}{2017}), \bibinfo{pages}{937--950}.
\newblock
\href{https://doi.org/10.1109/TPAMI.2016.2560814}{doi:\nolinkurl{10.1109/TPAMI.2016.2560814}}


\bibitem[Liu et~al\mbox{.}(2005)]%
        {Liu:2005}
\bibfield{author}{\bibinfo{person}{Ce Liu}, \bibinfo{person}{Antonio Torralba},
  \bibinfo{person}{William~T. Freeman}, \bibinfo{person}{Fr\'{e}do Durand},
  {and} \bibinfo{person}{Edward~H. Adelson}.} \bibinfo{year}{2005}\natexlab{}.
\newblock \showarticletitle{Motion magnification}.
\newblock \bibinfo{journal}{\emph{ACM Trans. Graph.}}  \bibinfo{volume}{24}
  (\bibinfo{date}{Jul} \bibinfo{year}{2005}), \bibinfo{pages}{519--526}.
\newblock
Issue 3.
\showISSN{0730-0301}
\href{https://doi.org/10.1145/1073204.1073223}{doi:\nolinkurl{10.1145/1073204.1073223}}


\bibitem[Luk{\'a}{\v{s}} et~al\mbox{.}(2006)]%
        {lukavs2006detecting}
\bibfield{author}{\bibinfo{person}{Jan Luk{\'a}{\v{s}}},
  \bibinfo{person}{Jessica Fridrich}, {and} \bibinfo{person}{Miroslav Goljan}.}
  \bibinfo{year}{2006}\natexlab{}.
\newblock \showarticletitle{Detecting digital image forgeries using sensor
  pattern noise}. In \bibinfo{booktitle}{\emph{Security, Steganography, and
  Watermarking of Multimedia Contents VIII}}, Vol.~\bibinfo{volume}{6072}.
  International Society for Optics and Photonics, \bibinfo{pages}{60720Y}.
\newblock


\bibitem[Ma et~al\mbox{.}(2022)]%
        {totems}
\bibfield{author}{\bibinfo{person}{Jingwei Ma}, \bibinfo{person}{Lucy Chai},
  \bibinfo{person}{Minyoung Huh}, \bibinfo{person}{Tongzhou Wang},
  \bibinfo{person}{Ser-Nam Lim}, \bibinfo{person}{Phillip Isola}, {and}
  \bibinfo{person}{Antonio Torralba}.} \bibinfo{year}{2022}\natexlab{}.
\newblock \showarticletitle{Totems: Physical Objects for Verifying Visual
  Integrity}. In \bibinfo{booktitle}{\emph{Computer Vision – ECCV 2022: 17th
  European Conference, Tel Aviv, Israel, October 23–27, 2022, Proceedings,
  Part XIV}} (Tel Aviv, Israel). \bibinfo{publisher}{Springer-Verlag},
  \bibinfo{address}{Berlin, Heidelberg}, \bibinfo{pages}{164–180}.
\newblock
\showISBNx{978-3-031-19780-2}
\href{https://doi.org/10.1007/978-3-031-19781-9_10}{doi:\nolinkurl{10.1007/978-3-031-19781-9_10}}


\bibitem[Michael Brice-Saddler(2019)]%
        {feinstein}
\bibfield{author}{\bibinfo{person}{The Washington~Post Michael Brice-Saddler}.}
  \bibinfo{year}{2019}\natexlab{}.
\newblock \bibinfo{booktitle}{\emph{Schoolchildren debate Dianne Feinstein on
  ‘Green New Deal.’ Her reply? ‘I know what I’m doing.’}}.
\newblock
\urldef\tempurl%
\url{https://www.washingtonpost.com/politics/2019/02/23/schoolchildren-debate-dianne-feinstein-green-new-deal-her-reply-i-know-what-im-doing/}
\showURL{%
\tempurl}


\bibitem[Mohanarathinam et~al\mbox{.}(2020)]%
        {watermarkreview}
\bibfield{author}{\bibinfo{person}{A. Mohanarathinam}, \bibinfo{person}{S.
  Kamalraj}, \bibinfo{person}{G.~K.~D. Prasanna~Venkatesan},
  \bibinfo{person}{Renjith~V. Ravi}, {and} \bibinfo{person}{C.~S.
  Manikandababu}.} \bibinfo{year}{2020}\natexlab{}.
\newblock \showarticletitle{Digital watermarking techniques for image security:
  a review}.
\newblock \bibinfo{journal}{\emph{Journal of Ambient Intelligence and Humanized
  Computing}} \bibinfo{volume}{11}, \bibinfo{number}{8} (\bibinfo{year}{2020}),
  \bibinfo{pages}{3221--3229}.
\newblock
\showISBNx{1868-5145}
\href{https://doi.org/10.1007/s12652-019-01500-1}{doi:\nolinkurl{10.1007/s12652-019-01500-1}}


\bibitem[Moreno et~al\mbox{.}(2015)]%
        {unstructuredlight}
\bibfield{author}{\bibinfo{person}{Daniel Moreno}, \bibinfo{person}{Fatih
  Calakli}, {and} \bibinfo{person}{Gabriel Taubin}.}
  \bibinfo{year}{2015}\natexlab{}.
\newblock \showarticletitle{Unsynchronized structured light}.
\newblock \bibinfo{journal}{\emph{ACM Trans. Graph.}} \bibinfo{volume}{34},
  \bibinfo{number}{6}, Article \bibinfo{articleno}{178} (\bibinfo{date}{nov}
  \bibinfo{year}{2015}), \bibinfo{numpages}{11}~pages.
\newblock
\showISSN{0730-0301}
\href{https://doi.org/10.1145/2816795.2818062}{doi:\nolinkurl{10.1145/2816795.2818062}}


\bibitem[Murmann et~al\mbox{.}(2016)]%
        {bounceflash}
\bibfield{author}{\bibinfo{person}{Lukas Murmann}, \bibinfo{person}{Abe Davis},
  \bibinfo{person}{Jan Kautz}, {and} \bibinfo{person}{Fr\'{e}do Durand}.}
  \bibinfo{year}{2016}\natexlab{}.
\newblock \showarticletitle{Computational bounce flash for indoor portraits}.
\newblock \bibinfo{journal}{\emph{ACM Trans. Graph.}} \bibinfo{volume}{35},
  \bibinfo{number}{6}, Article \bibinfo{articleno}{190} (\bibinfo{date}{dec}
  \bibinfo{year}{2016}), \bibinfo{numpages}{9}~pages.
\newblock
\showISSN{0730-0301}
\href{https://doi.org/10.1145/2980179.2980219}{doi:\nolinkurl{10.1145/2980179.2980219}}


\bibitem[Neekhara et~al\mbox{.}(2024)]%
        {facesigns}
\bibfield{author}{\bibinfo{person}{Paarth Neekhara}, \bibinfo{person}{Shehzeen
  Hussain}, \bibinfo{person}{Xinqiao Zhang}, \bibinfo{person}{Ke Huang},
  \bibinfo{person}{Julian McAuley}, {and} \bibinfo{person}{Farinaz
  Koushanfar}.} \bibinfo{year}{2024}\natexlab{}.
\newblock \showarticletitle{FaceSigns: Semi-fragile Watermarks for Media
  Authentication}.
\newblock \bibinfo{journal}{\emph{ACM Trans. Multimedia Comput. Commun. Appl.}}
  \bibinfo{volume}{20}, \bibinfo{number}{11}, Article \bibinfo{articleno}{337}
  (\bibinfo{date}{Sept.} \bibinfo{year}{2024}), \bibinfo{numpages}{21}~pages.
\newblock
\showISSN{1551-6857}
\href{https://doi.org/10.1145/3640466}{doi:\nolinkurl{10.1145/3640466}}


\bibitem[{NEMA}(2022)]%
        {ansi2022dimming}
\bibfield{author}{\bibinfo{person}{{NEMA}}.} \bibinfo{year}{2022}\natexlab{}.
\newblock \showarticletitle{0-10V Dimming Interface For LED Drivers,
  Fluorescent Ballasts, And Controls}.
\newblock  \bibinfo{number}{ANSI C137.1-2022} (\bibinfo{year}{2022}).
\newblock
\urldef\tempurl%
\url{https://www.nema.org/standards/view/american-national-standard-for-lighting-systems-0-10v-dimming-interface-for-led-drivers-fluorescent-ballasts-and-control}
\showURL{%
\tempurl}


\bibitem[Nikolaidis and Pitas(1999)]%
        {imagewatermark}
\bibfield{author}{\bibinfo{person}{N. Nikolaidis} {and} \bibinfo{person}{I.
  Pitas}.} \bibinfo{year}{1999}\natexlab{}.
\newblock \showarticletitle{Digital image watermarking: an overview}. In
  \bibinfo{booktitle}{\emph{Proceedings IEEE International Conference on
  Multimedia Computing and Systems}}, Vol.~\bibinfo{volume}{1}.
  \bibinfo{pages}{1--6 vol.1}.
\newblock
\href{https://doi.org/10.1109/MMCS.1999.779111}{doi:\nolinkurl{10.1109/MMCS.1999.779111}}


\bibitem[O'Brien and Farid(2012)]%
        {faridreflections}
\bibfield{author}{\bibinfo{person}{James~F. O'Brien} {and}
  \bibinfo{person}{Hany Farid}.} \bibinfo{year}{2012}\natexlab{}.
\newblock \showarticletitle{Exposing photo manipulation with inconsistent
  reflections}.
\newblock \bibinfo{journal}{\emph{ACM Trans. Graph.}} \bibinfo{volume}{31},
  \bibinfo{number}{1}, Article \bibinfo{articleno}{4} (\bibinfo{date}{feb}
  \bibinfo{year}{2012}), \bibinfo{numpages}{11}~pages.
\newblock
\showISSN{0730-0301}
\href{https://doi.org/10.1145/2077341.2077345}{doi:\nolinkurl{10.1145/2077341.2077345}}


\bibitem[Park et~al\mbox{.}(2007)]%
        {MultiSpectralMultiplexing}
\bibfield{author}{\bibinfo{person}{J. Park}, \bibinfo{person}{M. Lee},
  \bibinfo{person}{M.~D. Grossberg}, {and} \bibinfo{person}{S.~K. Nayar}.}
  \bibinfo{year}{2007}\natexlab{}.
\newblock \showarticletitle{{M}ultispectral {I}maging {U}sing {M}ultiplexed
  {I}llumination}. In \bibinfo{booktitle}{\emph{IEEE International Conference
  on Computer Vision (ICCV)}}.
\newblock


\bibitem[Parker~Molloy(2019)]%
        {bidenphysical}
\bibfield{author}{\bibinfo{person}{Media Matters for~America Parker~Molloy}.}
  \bibinfo{year}{2019}\natexlab{}.
\newblock \bibinfo{booktitle}{\emph{Right-wing media push false narrative that
  Biden called for a “Physical revolution”}}.
\newblock
\urldef\tempurl%
\url{https://www.mediamatters.org/mark-levin/right-wing-media-push-false-narrative-biden-called-physical-revolution}
\showURL{%
\tempurl}


\bibitem[Petschnigg et~al\mbox{.}(2004)]%
        {flashnoflash}
\bibfield{author}{\bibinfo{person}{Georg Petschnigg}, \bibinfo{person}{Richard
  Szeliski}, \bibinfo{person}{Maneesh Agrawala}, \bibinfo{person}{Michael
  Cohen}, \bibinfo{person}{Hugues Hoppe}, {and} \bibinfo{person}{Kentaro
  Toyama}.} \bibinfo{year}{2004}\natexlab{}.
\newblock \showarticletitle{Digital photography with flash and no-flash image
  pairs}.
\newblock \bibinfo{journal}{\emph{ACM Trans. Graph.}} \bibinfo{volume}{23},
  \bibinfo{number}{3} (\bibinfo{date}{aug} \bibinfo{year}{2004}),
  \bibinfo{pages}{664–672}.
\newblock
\showISSN{0730-0301}
\href{https://doi.org/10.1145/1015706.1015777}{doi:\nolinkurl{10.1145/1015706.1015777}}


\bibitem[Reuters(2020)]%
        {BidenShallowFake}
\bibfield{author}{\bibinfo{person}{Reuters}.} \bibinfo{year}{2020}\natexlab{}.
\newblock \bibinfo{booktitle}{\emph{Fact check: Video does not show Biden
  saying ‘Hello Minnesota’ in Florida rally}}.
\newblock
\urldef\tempurl%
\url{https://www.reuters.com/article/uk-factcheck-altered-sign-biden-mn/fact-check-video-does-not-show-biden-saying-hello-minnesota-in-florida-rally-idUSKBN27H1RZ}
\showURL{%
\tempurl}


\bibitem[Roberts(2013)]%
        {roberts2013undersampled}
\bibfield{author}{\bibinfo{person}{Richard~D Roberts}.}
  \bibinfo{year}{2013}\natexlab{}.
\newblock \showarticletitle{Undersampled frequency shift ON-OFF keying (UFSOOK)
  for camera communications (CamCom)}. In \bibinfo{booktitle}{\emph{2013 22nd
  Wireless and Optical Communication Conference}}. IEEE,
  \bibinfo{pages}{645--648}.
\newblock


\bibitem[Rusinkiewicz et~al\mbox{.}(2002)]%
        {structlight}
\bibfield{author}{\bibinfo{person}{Szymon Rusinkiewicz}, \bibinfo{person}{Olaf
  Hall-Holt}, {and} \bibinfo{person}{Marc Levoy}.}
  \bibinfo{year}{2002}\natexlab{}.
\newblock \showarticletitle{Real-time 3D model acquisition}.
\newblock \bibinfo{journal}{\emph{ACM Transactions on Graphics (TOG)}}
  \bibinfo{volume}{21}, \bibinfo{number}{3} (\bibinfo{year}{2002}),
  \bibinfo{pages}{438--446}.
\newblock


\bibitem[Sengupta et~al\mbox{.}(2021)]%
        {lightdesk}
\bibfield{author}{\bibinfo{person}{Soumyadip Sengupta}, \bibinfo{person}{Brian
  Curless}, \bibinfo{person}{Ira Kemelmacher-Shlizerman}, {and}
  \bibinfo{person}{Steven~M Seitz}.} \bibinfo{year}{2021}\natexlab{}.
\newblock \showarticletitle{A light stage on every desk}. In
  \bibinfo{booktitle}{\emph{Proceedings of the IEEE/CVF International
  Conference on Computer Vision}}. \bibinfo{pages}{2420--2429}.
\newblock


\bibitem[{Sheinin} et~al\mbox{.}(2017)]%
        {ImagingElectricGrid}
\bibfield{author}{\bibinfo{person}{M. {Sheinin}}, \bibinfo{person}{Y.~Y.
  {Schechner}}, {and} \bibinfo{person}{K.~N. {Kutulakos}}.}
  \bibinfo{year}{2017}\natexlab{}.
\newblock \showarticletitle{Computational Imaging on the Electric Grid}. In
  \bibinfo{booktitle}{\emph{2017 IEEE Conference on Computer Vision and Pattern
  Recognition (CVPR)}}. \bibinfo{pages}{2363--2372}.
\newblock
\href{https://doi.org/10.1109/CVPR.2017.254}{doi:\nolinkurl{10.1109/CVPR.2017.254}}


\bibitem[{Su} et~al\mbox{.}(2018)]%
        {8578766}
\bibfield{author}{\bibinfo{person}{S. {Su}}, \bibinfo{person}{F. {Heide}},
  \bibinfo{person}{G. {Wetzstein}}, {and} \bibinfo{person}{W. {Heidrich}}.}
  \bibinfo{year}{2018}\natexlab{}.
\newblock \showarticletitle{Deep End-to-End Time-of-Flight Imaging}. In
  \bibinfo{booktitle}{\emph{2018 IEEE/CVF Conference on Computer Vision and
  Pattern Recognition}}. \bibinfo{pages}{6383--6392}.
\newblock
\href{https://doi.org/10.1109/CVPR.2018.00668}{doi:\nolinkurl{10.1109/CVPR.2018.00668}}


\bibitem[Tominaga and Horiuchi(2012)]%
        {opticaspectral}
\bibfield{author}{\bibinfo{person}{Shoji Tominaga} {and}
  \bibinfo{person}{Takahiko Horiuchi}.} \bibinfo{year}{2012}\natexlab{}.
\newblock \showarticletitle{Spectral imaging by synchronizing capture and
  illumination}.
\newblock \bibinfo{journal}{\emph{J. Opt. Soc. Am. A}} \bibinfo{volume}{29},
  \bibinfo{number}{9} (\bibinfo{date}{Sep} \bibinfo{year}{2012}),
  \bibinfo{pages}{1764--1775}.
\newblock
\href{https://doi.org/10.1364/JOSAA.29.001764}{doi:\nolinkurl{10.1364/JOSAA.29.001764}}


\bibitem[Velten et~al\mbox{.}(2012)]%
        {Velten12recoveringthreedimensional}
\bibfield{author}{\bibinfo{person}{Andreas Velten}, \bibinfo{person}{Thomas
  Willwacher}, \bibinfo{person}{Otkrist Gupta}, \bibinfo{person}{Ashok
  Veeraraghavan}, \bibinfo{person}{Moungi~G. Bawendi}, {and}
  \bibinfo{person}{Ramesh Raskar}.} \bibinfo{year}{2012}\natexlab{}.
\newblock \showarticletitle{Recovering three-dimensional shape around a corner
  using ultrafast time-of-flight imaging}.
\newblock \bibinfo{journal}{\emph{Nature}} \bibinfo{volume}{3},
  \bibinfo{number}{1} (\bibinfo{year}{2012}), \bibinfo{pages}{745}.
\newblock


\bibitem[Wadhwa et~al\mbox{.}(2013)]%
        {PhaseBasedMM13}
\bibfield{author}{\bibinfo{person}{Neal Wadhwa}, \bibinfo{person}{Michael
  Rubinstein}, \bibinfo{person}{Fr\'{e}do Durand}, {and}
  \bibinfo{person}{William~T. Freeman}.} \bibinfo{year}{2013}\natexlab{}.
\newblock \showarticletitle{Phase-Based Video Motion Processing}.
\newblock \bibinfo{journal}{\emph{ACM Trans. Graph. (Proceedings SIGGRAPH
  2013)}} \bibinfo{volume}{32}, \bibinfo{number}{4} (\bibinfo{year}{2013}),
  \bibinfo{pages}{1--10}.
\newblock


\bibitem[Wang et~al\mbox{.}(2019)]%
        {wang2019detecting}
\bibfield{author}{\bibinfo{person}{Sheng-Yu Wang}, \bibinfo{person}{Oliver
  Wang}, \bibinfo{person}{Andrew Owens}, \bibinfo{person}{Richard Zhang}, {and}
  \bibinfo{person}{Alexei~A. Efros}.} \bibinfo{year}{2019}\natexlab{}.
\newblock \showarticletitle{Detecting Photoshopped Faces by Scripting
  Photoshop}. In \bibinfo{booktitle}{\emph{Proceedings of the IEEE/CVF
  International Conference on Computer Vision (ICCV)}}.
\newblock


\bibitem[Wang et~al\mbox{.}(2023)]%
        {freq}
\bibfield{author}{\bibinfo{person}{Yuan Wang}, \bibinfo{person}{Kun Yu},
  \bibinfo{person}{Chen Chen}, \bibinfo{person}{Xiyuan Hu}, {and}
  \bibinfo{person}{Silong Peng}.} \bibinfo{year}{2023}\natexlab{}.
\newblock \showarticletitle{Dynamic Graph Learning With Content-Guided
  Spatial-Frequency Relation Reasoning for Deepfake Detection}. In
  \bibinfo{booktitle}{\emph{Proceedings of the IEEE/CVF Conference on Computer
  Vision and Pattern Recognition (CVPR)}}. \bibinfo{pages}{7278--7287}.
\newblock


\bibitem[Wenger et~al\mbox{.}(2005)]%
        {lightstage}
\bibfield{author}{\bibinfo{person}{Andreas Wenger}, \bibinfo{person}{Andrew
  Gardner}, \bibinfo{person}{Chris Tchou}, \bibinfo{person}{Jonas Unger},
  \bibinfo{person}{Tim Hawkins}, {and} \bibinfo{person}{Paul Debevec}.}
  \bibinfo{year}{2005}\natexlab{}.
\newblock \showarticletitle{Performance Relighting and Reflectance
  Transformation with Time-Multiplexed Illumination}.
\newblock \bibinfo{journal}{\emph{ACM Trans. Graph.}} \bibinfo{volume}{24},
  \bibinfo{number}{3} (\bibinfo{date}{jul} \bibinfo{year}{2005}),
  \bibinfo{pages}{756–764}.
\newblock
\showISSN{0730-0301}
\href{https://doi.org/10.1145/1073204.1073258}{doi:\nolinkurl{10.1145/1073204.1073258}}


\bibitem[Wu et~al\mbox{.}(2012)]%
        {wu2012eulerian}
\bibfield{author}{\bibinfo{person}{Hao-Yu Wu}, \bibinfo{person}{Michael
  Rubinstein}, \bibinfo{person}{Eugene Shih}, \bibinfo{person}{John Guttag},
  \bibinfo{person}{Fr{\'e}do Durand}, {and} \bibinfo{person}{William Freeman}.}
  \bibinfo{year}{2012}\natexlab{}.
\newblock \showarticletitle{Eulerian video magnification for revealing subtle
  changes in the world}.
\newblock \bibinfo{journal}{\emph{ACM Transactions on Graphics (TOG)}}
  \bibinfo{volume}{31}, \bibinfo{number}{4} (\bibinfo{year}{2012}),
  \bibinfo{pages}{1--8}.
\newblock


\bibitem[Xiang et~al\mbox{.}(2023)]%
        {mtn}
\bibfield{author}{\bibinfo{person}{Ziyue Xiang}, \bibinfo{person}{Amit
  Kumar~Singh Yadav}, \bibinfo{person}{Paolo Bestagini},
  \bibinfo{person}{Stefano Tubaro}, {and} \bibinfo{person}{Edward~J. Delp}.}
  \bibinfo{year}{2023}\natexlab{}.
\newblock \showarticletitle{MTN: Forensic Analysis of MP4 Video Files Using
  Graph Neural Networks}. In \bibinfo{booktitle}{\emph{Proceedings of the
  IEEE/CVF Conference on Computer Vision and Pattern Recognition (CVPR)
  Workshops}}. \bibinfo{pages}{963--972}.
\newblock


\bibitem[Ye et~al\mbox{.}(2019)]%
        {spi}
\bibfield{author}{\bibinfo{person}{Zhiyuan Ye}, \bibinfo{person}{Panghe Qiu},
  \bibinfo{person}{Haibo Wang}, \bibinfo{person}{Jun Xiong}, {and}
  \bibinfo{person}{Kaige Wang}.} \bibinfo{year}{2019}\natexlab{}.
\newblock \showarticletitle{Image watermarking and fusion based on Fourier
  single-pixel imaging with weighed light source}.
\newblock \bibinfo{journal}{\emph{Opt. Express}} \bibinfo{volume}{27},
  \bibinfo{number}{25} (\bibinfo{date}{Dec} \bibinfo{year}{2019}),
  \bibinfo{pages}{36505--36523}.
\newblock
\href{https://doi.org/10.1364/OE.27.036505}{doi:\nolinkurl{10.1364/OE.27.036505}}


\bibitem[Zhao et~al\mbox{.}(2023)]%
        {proactive}
\bibfield{author}{\bibinfo{person}{Yuan Zhao}, \bibinfo{person}{Bo Liu},
  \bibinfo{person}{Ming Ding}, \bibinfo{person}{Baoping Liu},
  \bibinfo{person}{Tianqing Zhu}, {and} \bibinfo{person}{Xin Yu}.}
  \bibinfo{year}{2023}\natexlab{}.
\newblock \showarticletitle{Proactive deepfake defence via identity
  watermarking}. In \bibinfo{booktitle}{\emph{Proceedings of the IEEE/CVF
  winter conference on applications of computer vision}}.
  \bibinfo{pages}{4602--4611}.
\newblock


\end{thebibliography}

\appendix

\begin{figure}[ht]
     \centering
     \includegraphics[width=0.7\linewidth]{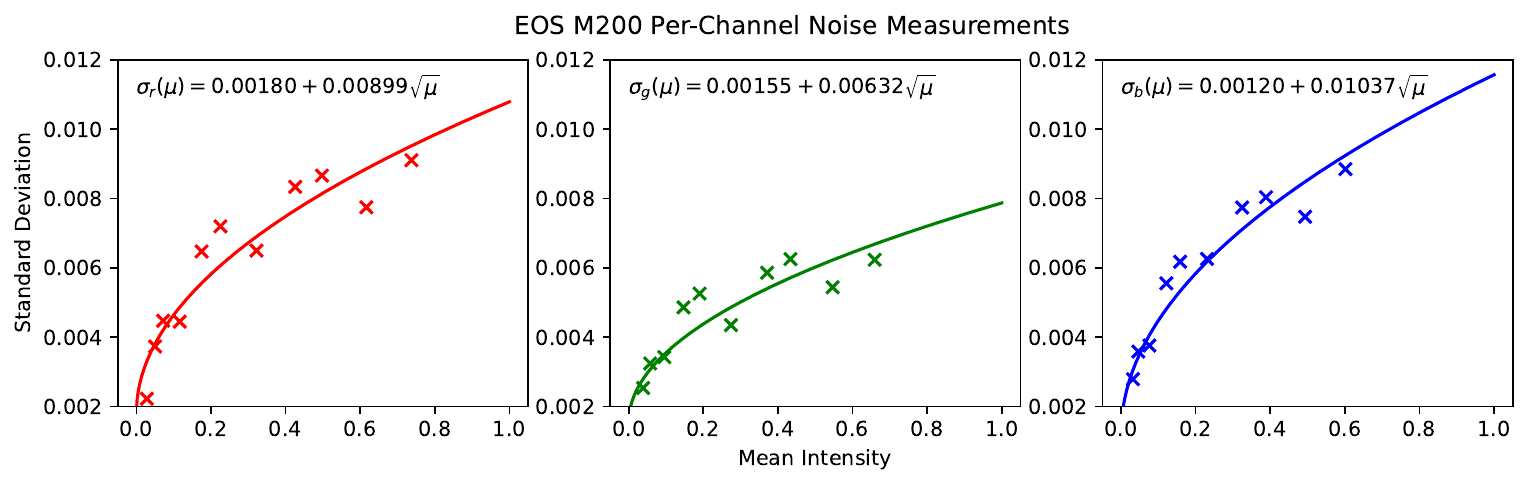}
     \includegraphics[width=0.2\linewidth]{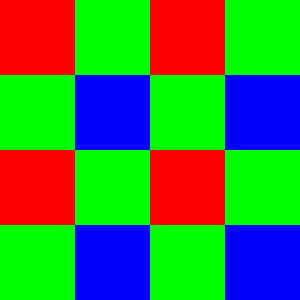}
     \caption{\textbf{Per-Channel Read and Photon Noise Empirical Trend} From left to right we get the least squares empirical estimate of the read and photon noise for the red, green, and blue channels respectively. Note how the green channel has less noise likely due to the Bayer pattern, shown on the right. Note that the constant term corresponds to read noise.}
     \Description{Left: empirical read and photon noise trends for the red, green, and blue channels. Right: Vector drawing of Bayer pattern for the camera sensor.}
     \label{fig:photon}
    %\caption{\textbf{SNR with Photon and Read Noise Sources}}
\end{figure}

\begin{figure}[ht]
        \centering
        \includegraphics[width=0.5\linewidth]{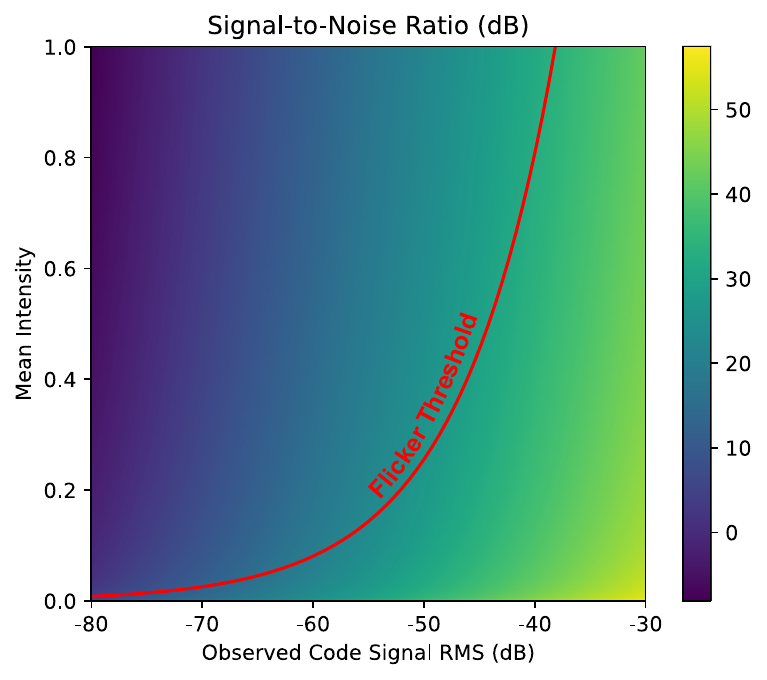}
     \caption{\textbf{Signal-to-Noise Ratio Simulation} The above plot shows simulated signal-to-noise ratios for different signal levels (both in decibels) and mean brightnesses assuming a window size of 450 frames and a downsampling factor of 2, used for most experiments in our paper. This uses our empirical noise variance measurements (red channel). We use the code signal from the dark-skin subject example in the supplementary to calculate our (distracted) flicker detection threshold. Note that increasing the signal level increases SNR while increasing mean brightness decreases SNR due to photon noise. Additionally, as the mean intensity increases, we become less sensitive to flicker, allowing for the use of a higher signal level which results in higher SNR. Staying left of the curve keeps us under the flicker threshold.}
     \Description{Plot of simulated signal-to-noise ratio for different signal levels and mean brightnesses with a superimposed trendline for the theoretical flicker sensitivty threshold.}
     \label{fig:photon_snr}
\end{figure}

\section{Predicting SNR}
\label{sec:snr}
Our supplemental material includes analyses of several capture conditions and noise sources (e.g., photon noise and compression) that contribute to the SNR of \nci{}. Here, we summarize some of these findings in an analytic expression for SNR parameterized by code signal strength, the amount of uncoded light $L$, temporal analysis window size $w$, and spatial downsample factor $M$ to help predict the behavior of \nci{} in different settings under photon and read noise.
Specifically, we consider the SNR associated with recovering transfer coefficients. Our signal is given by the expectation in \Eq{} \eqref{eq:singlepixcodeimage}:
\begin{align}
\text{signal} &= \E\left[
\frac{\codew{}^\intercal \obsw{}}{\codew{}^\intercal\codew{}}
\right] = \reflectance{}
\end{align}
and our noise is given by the standard deviation, which only depends on the noise portion of $\obsw{}$,
\begin{align}
\text{noise} &= 
\V\left[
\frac{\codew{}^\intercal \obsw{}}{\codew{}^\intercal\codew{}}
\right]=\V\left[
\frac{\codew{}^\intercal \noisew{}}{\codew{}^\intercal\codew{}}
\right]
= \frac{\stdn{}}{||\codew{}||_2}
\end{align}

% \begin{align}
% \text{signal} &= \E\left[
% \frac{\codew{}^\intercal \obsw{}}{\codew{}^\intercal\codew{}}
% \right],  &
% \text{noise} &= 
% \V\left[
% \frac{\codew{}^\intercal \obsw{}}{\codew{}^\intercal\codew{}}
% \right]
% \end{align}

% We will consider the tranRecall the expectation of our code image equation for a single pixel and coded source:
% \beq
% \E\left[
% \frac{\codew{}^\intercal \obsw{}}{\codew{}^\intercal\codew{}}
% \right]=\reflectance{}
% \label{eq:singlepixcodeimage}
% \eeq
% which represents our recovered signal. The standard deviation of this equation represents the strength of our noise, which depends only on $\noisew{}$:

% \noindent We consider code signal strength, 
% % % $\Rms\left[\codew{}\right]r$, 
% % $\codestrength{}$ where $\codew{}=\codestrength{}\normcodew{}$
% the mean brightness of uncoded light $L$, temporal analysis window size $w$, and spatial downsample factor $\spatial$. We consider photon and read noise. 

% Recall the expectation of our code image equation for a single pixel and coded source:
% \beq
% \E\left[
% \frac{\codew{}^\intercal \obsw{}}{\codew{}^\intercal\codew{}}
% \right]=\reflectance{}
% \label{eq:singlepixcodeimage}
% \eeq
% which represents our recovered signal. 
% The standard deviation of this equation represents the strength of our noise, which depends only on $\noisew{}$:

% \beq
% \V\left[
% \frac{\codew{}^\intercal \obsw{}}{\codew{}^\intercal\codew{}}
% \right] = \V\left[
% \frac{\codew{}^\intercal \noisew{}}{\codew{}^\intercal\codew{}}
% \right] = \frac{\stdn{}}{||\codew{}||_2}\\
% \eeq
\noindent We provide a step-by-step derivation of this equality in the supplemental. $\stdn{}$ is the standard deviation of our read and photon noise, where the latter scales proportionally to a square root factor of the uncoded brightness of the scene $L$. This term is approximately invariant to the code signal fluctuations since they are orders of magnitude smaller than the amount of uncoded light in a scene. We also know that our noise level should decrease when we spatially downsample a video, such as the extreme case where we take a framewise average for global temporal registration. By the law of large numbers, we know that the standard deviation decreases by the square root of the number of averaged samples $\spatial$. So if we downsample the height and width of a frame by a factor of 2, $\spatial=4$. We update our equation:

\beq
noise = \frac{\stdn{}}{||\codew{}||_2\sqrt{\spatial}}\\
\eeq

We further factor window size out of $||\codew{}||_2$ to better understand the individual effects of the code signal strength and window size $w$: 

\begin{align}
    noise &= \frac{\stdn{}}{||\codew{}||_2\sqrt{\spatial}}\\
    &= \frac{\stdn{}}{\frac{||\codew{}||_2}{\sqrt{w}}\sqrt{\spatial w}}\\
    &= \frac{\stdn{}}{\Rms\left[\codew{}\right]\sqrt{\spatial w}}\\
\end{align}

Where $\Rms$ is the root-mean-square. Our SNR is:
\begin{align}
SNR &= \frac{signal}{noise}\\
    &= \frac{r}{\frac{\stdn{}}{\Rms\left[\codew{}\right]\sqrt{\spatial w}}}\\
    &= \frac{\sqrt{\spatial w}\left(\Rms\left[\codew{}\right]r\right)}{\stdn{}}\\
    &= \frac{\sqrt{\spatial w}\left(\Rms\left[\codew{}\right]r\right)}{a + b\sqrt{L}}
\end{align}
Similar to the approach taken in \cite{lightstage}, we empirically measure $a$ and $b$ by capturing videos of a white wall with indoor lighting from a Canon EOS M200 camera under different apertures with a shutter speed of $\frac{1}{60}$s and constant ISO. We then take the mean and standard deviation of each pixel over time separately for each channel. We perform least squares square root regression per channel to get the result shown in Figure \ref{fig:photon}. Note how photon noise dominates read noise in most of the working range of the sensor except in low-light scenarios. Figure \ref{fig:photon_snr} shows a plot of a slice of this SNR function parameterized by our mean brightness, $L$, and our observed code signal RMS, $\Rms [\codew{}]\reflectance{}$.

\begin{table*}[ht]
\centering
\caption{Overview of Main Experimental Setups}
\Description{Left: Setup Name, Middle: Lighting/Recording Devices Used, Right: Image of Setup.}
\label{tab:setup}
\begin{tabular}{ |>{\centering\arraybackslash}m{2cm}||>{\centering\arraybackslash}m{4cm}|>{\centering\arraybackslash}m{10cm}|  }
 \hline
 \multicolumn{3}{|c|}{Experimental Setup List} \\
 \hline
 Setup Name & Devices Used& Setup Figure\\
 \hline
 Politician & \begin{itemize}[leftmargin=*] \item \textbf{Recording:} \begin{itemize} \item Canon EOS M200 \item iPhone XS (handheld) \end{itemize} \item \textbf{Light Sources: } \begin{itemize} \item Stage Light (coded) \item Room Lights (uncoded) \end{itemize}\end{itemize} & \includegraphics[width=\linewidth]{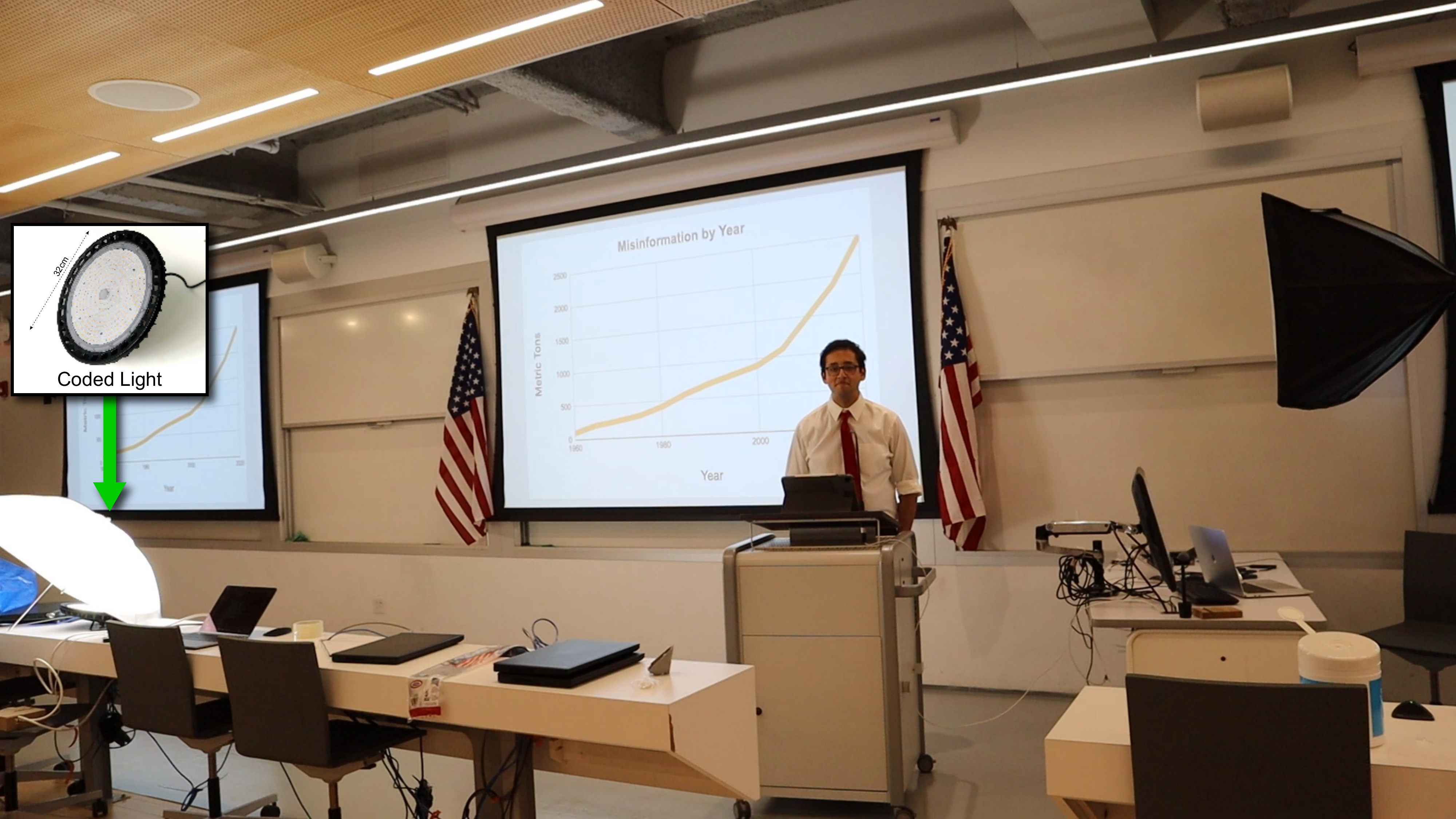}\\
 \hline
 \centering Teleconference & \begin{itemize}[leftmargin=*] \item \textbf{Recording:} \begin{itemize} \item Canon EOS M200\end{itemize} \item \textbf{Light Sources: } \begin{itemize} \item Computer Monitor (coded) \end{itemize}\end{itemize} & \includegraphics[width=\linewidth]{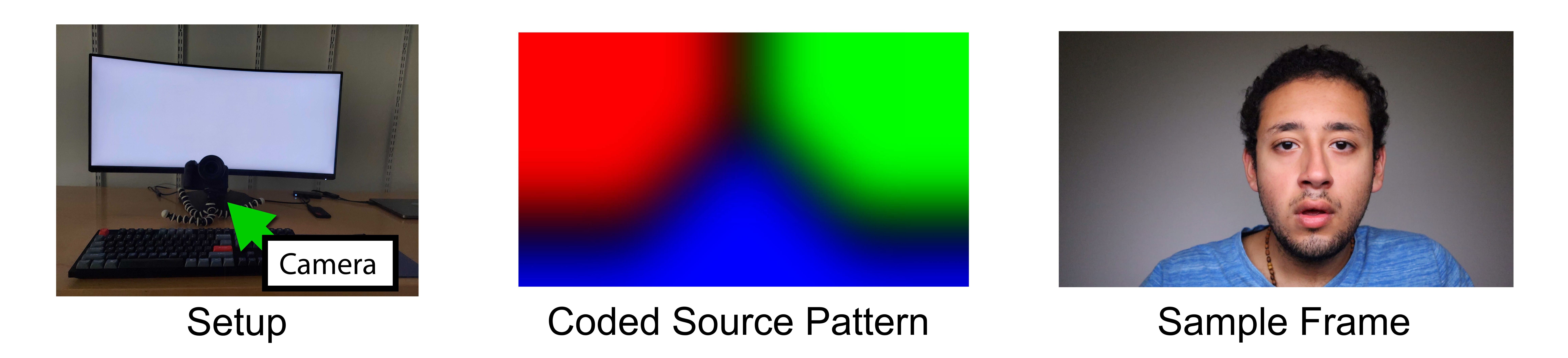}\\
 \hline
 Conference Presentation & \begin{itemize}[leftmargin=*] \item \textbf{Recording:} \begin{itemize} \item Canon EOS M200 \item Galaxy A71 5G (handheld) \end{itemize} \item \textbf{Light Sources: } \begin{itemize} \item Stage Light (coded) \item Right Television Screen (coded) \item Left Television Screen (uncoded) \item Room Lights (uncoded) \end{itemize}\end{itemize} & \includegraphics[width=\linewidth]{fig/experiments/caroline_setup.pdf}\\
 \hline
 Hackathon (actual event) & \begin{itemize}[leftmargin=*] \item \textbf{Recording:} \begin{itemize} \item Canon R6 Mark II \item iPhone 15 Pro Max (handheld) \end{itemize} \item \textbf{Light Sources: } \begin{itemize} \item Stage Light (coded) \item Room Lights (uncoded) \end{itemize}\end{itemize} & \includegraphics[width=\linewidth]{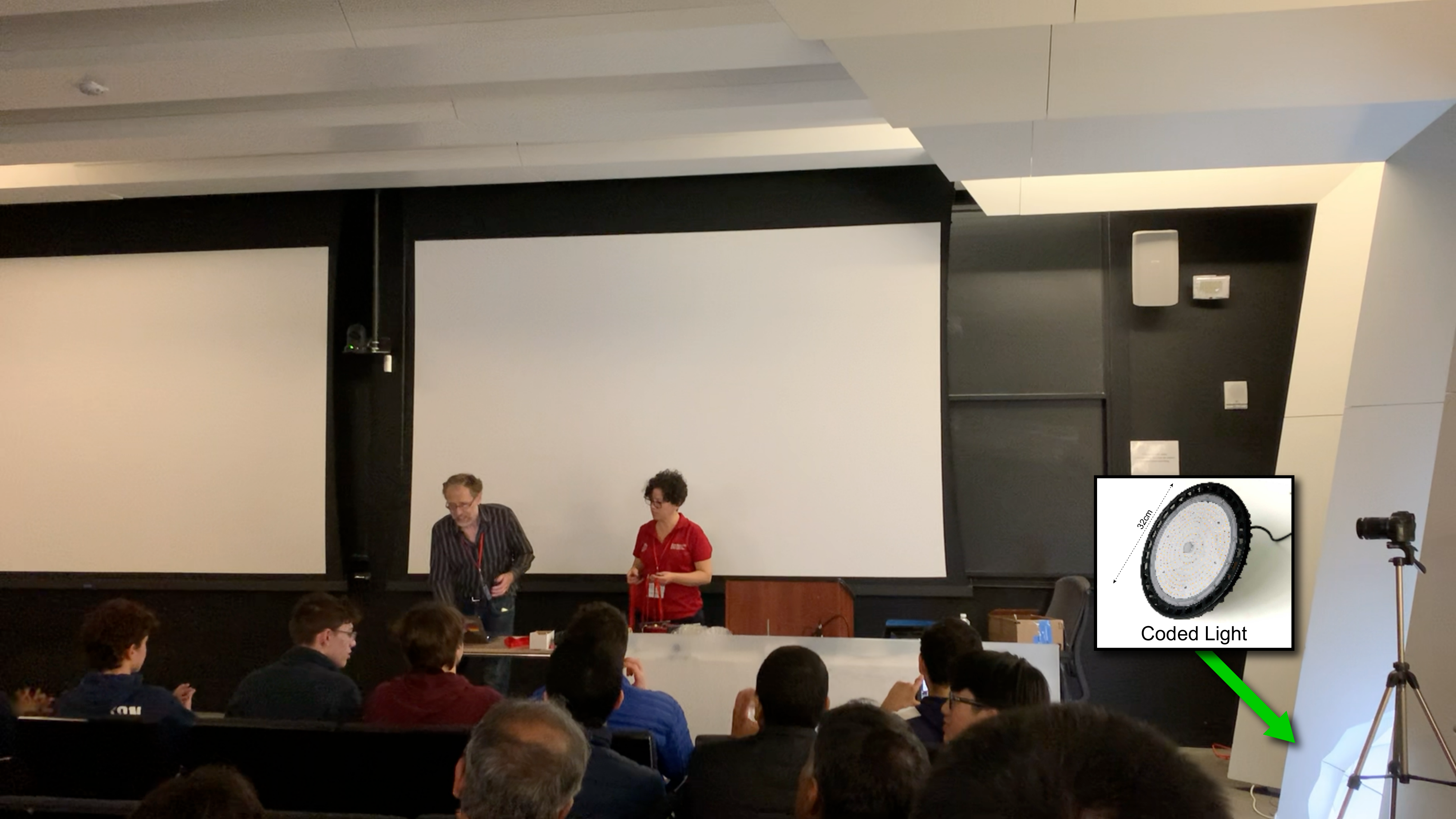}\\
 \hline
\end{tabular}
\end{table*}

\begin{figure*}[ht!]
    \centering
    \includegraphics[width=\linewidth]{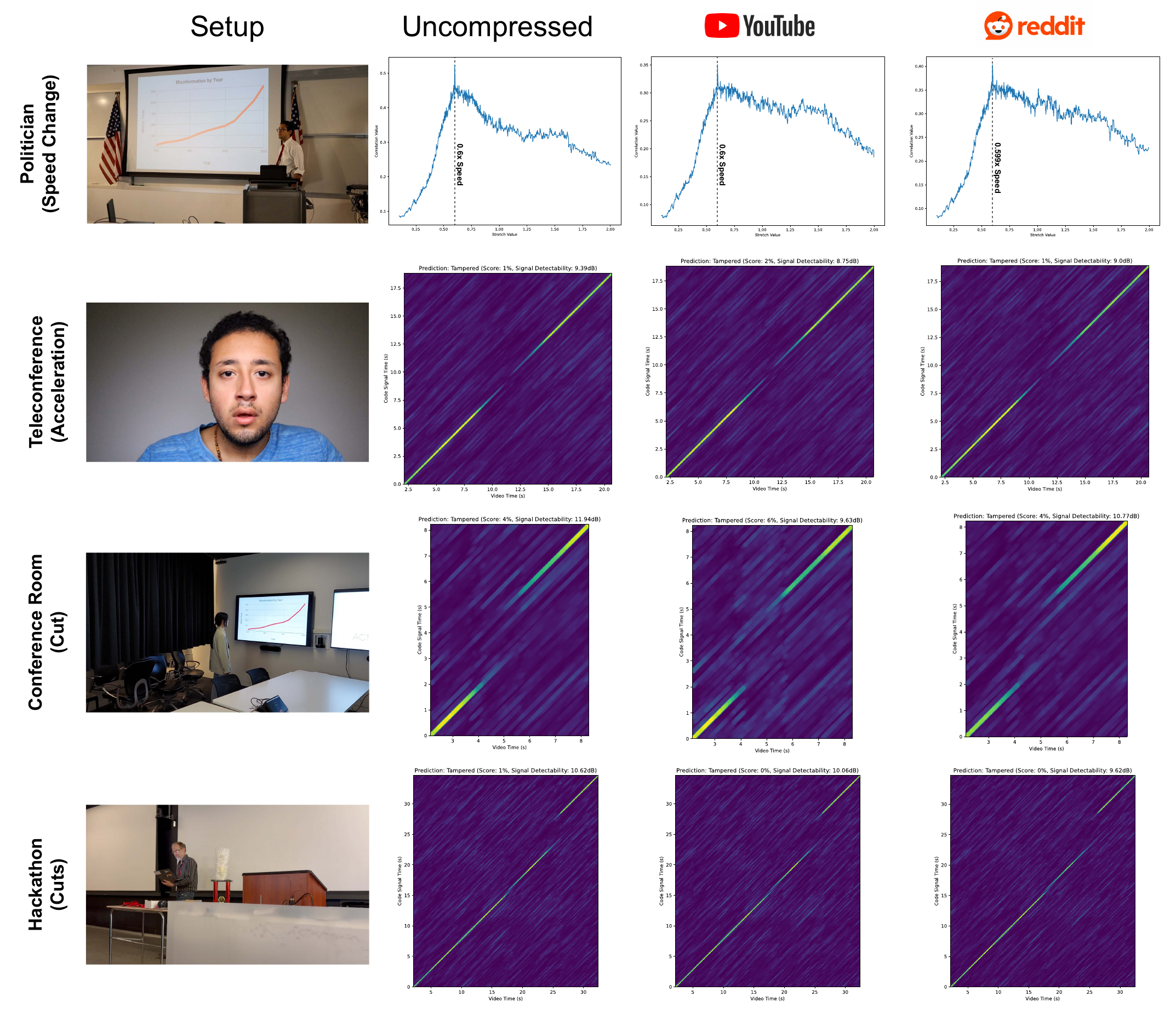}
    \caption{
    \textbf{Detecting Temporal Manipulation Under Compression from Popular Video Hosting Websites}
    We analyze \nci{} to detect temporal manipulation of four different scenes (rows) in videos with compression applied by different video hosting websites (columns). The video in the first row contains a global speed change, for which we visualize the likelihood of different speed changes (\Sectn{\ref{sec:characterizingspeedmanipulation}}). We correctly identify a 0.6x speed change in all cases. The video in the second row contains a localized acceleration, and the remaining two rows contain seamless temporal cuts implemented with Adobe's warp cut feature. In all cases, the time manipulation results in a visible discontinuity along the diagonal of the alignment matrix. Manipulated videos can be found in our supplemental material.
    % search over different  in the first row frequency-domain  results of our method for detecting a speed change ()
    % \textbf{Detecting Temporal Tampering} In the first example (column), we slowed down the handheld iPhone XS politician video.
    % similar to the tampered video of Nancy Pelosi that made it look like she had slurred speech \cite{PelosiSlur}. 
    % Our Fourier-domain search algorithm finds a peak at the true stretch factor of 0.6x. In the second example, the teleconferencing video (in front of a computer monitor) was sped up at around 11 seconds for about 3 seconds to make the person seem like they counted faster for a portion of time. In the third example, we cut out the word "not" from a video captured by a Samsung Galaxy A71 5G in the conference presentation setup to make the subject say that the problem of misinformation is solved. In the last example, we cut the word "effort" in the handheld iPhone 15 Pro hackathon video as well as a few words after to make it sound like the team won the "best award" instead of the "best effort award". The disconnection and misalignment of the otherwise linear time correspondence in the matrices indicate temporally tampered videos as shown by our automatic detection algorithm's results in the title of each plot. We use a window size of 4.3 seconds for the last three examples, and the whole video for the speed one, though we could do it on smaller chunks.}
    }
    \Description{First column: Frame from temporally manipulated video. Second column: temporal analysis visualizations for uncompressed manipulated video. Third column: visualizations for YouTube compressed manipulated video. Fourth column: visualizations for Reddit compressed manipulated video.}
    \label{fig:dtw}
\end{figure*}
\begin{figure*}[ht]
    \includegraphics[width=\linewidth]{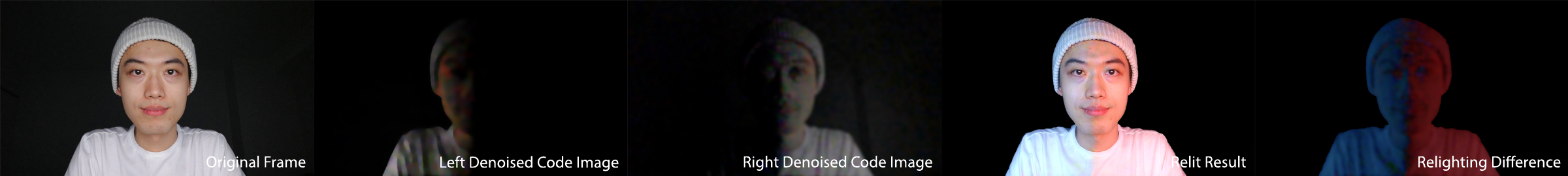}
    \caption{\textbf{Relighting (Similar to Teleconference Setup)} Code images can be used for relighting since they represent the scene lit by the respective coded source. After denoising the code images, they can be imported into Photoshop for relighting the target frame in the video.}
    \Description{Left to Right: video frame, denoised code image from half of monitor lighting left half of face, code image from half of monitor lighting right half of face, relit video frame, relighting difference image.}
    \label{fig:relit}
\end{figure*}
\end{document}